\documentstyle[aps,12pt,epsf]{revtex}
\begin{document}
\title{Multiple pion production from an oriented chiral condensate}
\author{Alexander Volya, Scott Pratt and Vladimir Zelevinsky}
\address{Department of Physics and Astronomy and \\
National Superconducting Cyclotron Laboratory, \\
Michigan State University,\\
East Lansing, Michigan 48824-1321, USA}
\date{\today}
\maketitle

\begin{abstract}
\baselineskip 14pt
We consider an ``oriented'' chiral condensate produced in the squeezed states
of the effective field theory with time- and space-dependent pion mass 
parameter. We discuss the general properties of the solution, identifying
condensate modes and determining the resulting pion distributions.
The implementation of  the dynamics in the form of sudden perturbation
allows us to look for exact solutions.  In the region of condensation,
the dramatic increase in pion production and charge 
fluctuations are demonstrated. 
\end{abstract}

\baselineskip 14pt
\section{Introduction}
Pions have always played an important role in heavy ion reactions.
With their relatively strong coupling to nuclei and  
very small mass, pions can be produced in large quantities and  
can carry important detectable information 
about the state of nuclear matter during the reaction.
Large fluctuations in the ratio of charged to neutral pions have been
observed in 
cosmic ray experiments \cite{cosmic} and have recently been a subject of many
discussions. The major interest
in this direction is motivated by an almost a decade old idea 
\cite{anselm88,anselm94,bjorken93} that at high enough energies 
the pion-sigma ground state symmetry breaking condensate can be destroyed.
In the latter process of sudden cooling there is a chance of the formation of
disoriented chiral domains, given that 
the terms responsible for the explicit breaking 
of chiral symmetry are small.  
The formation of this disoriented chiral 
condensate (DCC) is presumably responsible for the observed discrepancy in
the ratio of pion species. 
Further enthusiasm about these ideas is generated by 
the fact that it may be possible to achieve an environment capable of 
producing disoriented chiral domains  in relativistic heavy ion collisions. 
In that case the large fluctuations in pion types can be a signal of a DCC.

There exist a number of theoretical works investigating the 
production and development of DCCs in heavy ion collisions. 
The assumption of a random 
equiprobable orientation of the pionic isovector  
leads \cite{anselm94} to the probability of observing a
fraction of neutral pions $f\equiv N_{\pi^{0}}/(N_{\pi^{+}}+N_{\pi^{-}}+
N_{\pi^{0}})$ being 
\begin{equation}
{\cal P}(f)={1\over 2\sqrt{f}}\, .
\label{probability} 
\end{equation} 
This result is consistent with
the slowly varying classical pionic field as a solution of the
non-linear $\sigma$-model \cite{anselm97}. It can also be obtained 
as a limiting large $N$ distribution in the coherent single-mode 
pion production by an isoscalar
operator \cite{kowalski92,pratt94,huang96_2} 
$$\left ({\vec a}\,^\dagger\cdot{\vec a}\,^\dagger\right )^{N/2}\,=\,
\left ( a^{\dagger}_{x} a^{\dagger}_{x}
 + a^{\dagger}_{y} a^{\dagger}_{y} + a^{\dagger}_{z} 
a^{\dagger}_{z} \right )^{N/2} \, .$$ 
The formation and the nature 
of  domain structures is an important question in itself
\cite{huang96,randrup97}. In the case of many small domains or
in the absence of a DCC,  one would expect a Gaussian 
distribution $P(f)\,,$ following from the central limit theorem.
More complicated methods of quantum and classical field theory applied to
this problem with  
different assumptions of formation, evolution \cite{kluger95,cooper95} 
and dissociation of DCCs \cite{lampert96,asakwa95} 
lead to different results.  Pionic mode mixing and  
final state interaction may also greatly change the observed 
forms of these probability distributions  \cite{huang96_2,anselm97,camelia97}.

Most of the existing theoretical works 
consider  pion production from DCC domains 
formed in the dynamical process of symmetry restoration 
and 
later sudden relaxation into a non-symmetric vacuum which is some 
times called quench \cite{rajagopal93}. 
The word ``disoriented'' also describes the theoretical foundation
of the DCC approach which is based on introducing an effective
disoriented current-type term in the Lagrangian.
The advantage of this formalism is in the simplicity of the solution,
as a current-type interaction can always be solved with a coherent
state formalism \cite{horn71}. 
In contrast to that 
the linear sigma model, a well-established effective theory based
on chiral symmetry, naturally leads to a quadratic pion field term in 
the Lagrangian. 
The goal of the present work is to consider 
an ``oriented'' chiral condensate with a chirally
symmetric Lagrangian that to the lowest order is quadratic in the pion field.
Throughout all this paper we are going to deal with the wave equation for 
pions  in the form 
\begin{equation}
{\partial^2 \vec{\pi} 
\over \partial t^2} - \nabla^2 \vec{\pi} + m_{\rm eff}^2({\bf x},t)\, \vec{\pi} = 0
 \, .
\label{fieldequation}
\end{equation}
This is the equation of
motion deduced from a quadratic Lagrangian.
Equation (\ref{fieldequation}) contains an 
effective mass  $m_{\rm eff}({\bf x},t)$ 
which is due to a mean field  from  non-pionic degrees of freedom.  
This term produces a parametric
excitation of the pionic field, and, if $m^2_{\rm eff}<0$, 
may lead to amplification of
low momentum modes and the formation of a chiral condensate. 
The chiral condensate in this picture is  ``oriented''; 
as opposed to the usual picture of a DCC. 
The scalar mean field (effective mass) 
introduced here does not
break the symmetry properties of pion fields in any way. 
The quantum  problems formulated by 
equations of the form  
(\ref{fieldequation}) 
address a variety of  physical issues, 
and can be encountered in different areas of physics.
In general, such non-stationary quantum  problems  cannot be solved 
exactly. Even studies with approximate methods  like perturbation theory,
impulse approximations, adiabatic expansions and many others often require 
sophisticated approaches. We will address
those  rare occasions
when exact solutions may be obtained.  
The significance of the analytical solution should not be underestimated. 
The exact results could exhibit
the unperturbative features of the solutions, like phase transitions and 
condensates, as well as point  the way to a good 
approximate theory. 

The paper is structured as follows. 
We start with a short introduction to the
linear sigma model and show that  effectively it leads to 
Eq. (\ref{fieldequation}) for a pion field.
In the next section we discuss 
an exact solution of Eq. (\ref{fieldequation}).
First, we  show that from the classical solution a quantum solution can
be built exactly. Then, based on the general form of the evolution matrix,
we  obtain and 
discuss  possible forms of the multiplicity distribution.
The bulk of the paper is concentrated in Sec. \ref{sec_application}. 
where we  further analyze
the properties of the solutions to Eq. (\ref{fieldequation}). 
By considering  a simpler model of space-independent but variable-in-time
effective mass we clearly define the exponentially growing ``condensate''
momentum modes. Depending on the number and strength of the condensate modes
we obtain different distributions for a particle number and
distributions over pion species.  We emphasize the limits 
when Eq. (\ref{probability}) is recovered or when ${\cal P}(f)$ becomes
Gaussian.  In the following part 
of Sec. \ref{sec_application} we  consider a full
field equation with a time- and space-dependent mass parameter and address
the exactly solvable case of a mass parameter abruptly 
changing and returning to normal. We show that in this picture
the condensate modes are still identifiable and
they still have a characteristic momentum distribution.
Final summary and conclusions are given in  Sec. 
\ref{sec_conclusions}.

\section{The Linear sigma model}
\label{sec_LSM}
Below we  give a short review of
the linear sigma model, point out the origin of Eq. 
(\ref{fieldequation}) in the context of pion fields 
and deliberate the possible form of the 
effective mass term.
We assume the usual linear sigma model Lagrangian 
of the pion and sigma fields \cite{koch97} 
\begin{equation}
{\cal L}_{L.S.}= {1\over 2} \left (\partial_{\mu} {\vec \pi}\right ) \cdot 
\left ( \partial^{\mu} 
{\vec \pi}\right ) + {1\over 2} \left ( \partial_{\mu}\sigma \right )
\,\left (\partial^{\mu}\sigma\right )  - 
{\lambda\over 4} \left ( ({\vec\pi}^2 +\sigma^2)-v_{0}^2 \right )^2 + 
\epsilon \sigma  \, , 
\label{sigmamodel}
\end{equation}       
where explicit symmetry breaking is introduced by the parameter $\epsilon$. 
The sigma field has a non-zero vacuum expectation value $f_{\pi}$ 
that is set by the Goldberger-Treiman relation and is 
related to the above mentioned symmetry breaking parameter as
$$v_{0}^2=f_{\pi}^2 - {\epsilon\over \lambda f_{\pi}} \, .$$
The Lagrangian of Eq. (\ref{sigmamodel}) produces the following masses
of the pion and sigma mesons
$$ m_{\pi}^2={\epsilon\over f_{\pi}}\, , 
\qquad m_{\sigma}^2=2\lambda f_{\pi}^2 +
{\epsilon \over f_{\pi}} \, . 
$$
Equations of motion for each isospin component of the pion field $\pi_\tau$ 
have the form
\begin{equation}
{\partial^2 \pi_\tau \over \partial t^2} - \nabla^2 \pi_\tau + \lambda 
\left ( ({\vec \pi}^2+\sigma^2)-v_0^2 \right ) \pi_\tau = 0 \, .
\label{eqmotion}
\end{equation}
In the mean field approximation non-pionic  degrees of freedom,
like the sigma field, are approximated by their expectation value.
Then the dynamics 
of the pionic field can be viewed as a field equation 
of type (\ref{fieldequation}), 
with a time- and space-dependent 
mass parameter 
$$m_{\rm eff}^2 ({\bf x},t)= \lambda \left ( \langle 
{\vec \pi}^2+\sigma^2\rangle-v_0^2 \right ) \, .$$

As in the application to the 
chiral condensates, we would like to solve  
Eq. (\ref{eqmotion}) in the formalism of quantum field theory with
reasonable expectation values  
$\langle{\vec \pi}^2+\sigma^2\rangle$.
We do not solve the problem self-consistently because 
$m_{\rm eff}({\bf x},t)\, ,$ being generated by other degrees of freedom,
is placed in by hand as a  mean field.     
Before and long after the reaction  the fields are
in their ground state so that $\langle \pi_\tau \rangle = 0 $ and 
$\langle \sigma \rangle = f_{\pi}$ , which is equivalent to
$$m^2_{\rm eff}({\bf x},t=-\infty)=m^2_{\rm eff}({\bf x},t=+\infty) = 
m^2_{\pi} \,.$$
During the reaction, the behavior of the effective mass is unknown,
and in our case would be an input to the model. 
When chiral symmetry is restored the expectation of the $\sigma$ field
tends to zero, while the effective squared pion mass  is positive. 
A sudden return of the effective potential to its vacuum form (quenching)
can strand the $\sigma$ mean field near zero with a large negative 
value of the effective pion mass squared, which in the extreme limit would 
reach $ m_{\rm eff}^2 = -\lambda v_0^2 \approx -m_{\sigma}^2/2 \, .$
A negative value of $ m_{\rm eff}^2$  would lead to an exponential 
growth of the  low momentum pion modes and long range correlations, i.e. 
the creation of a chiral condensate \cite{rajagopal95}. 

\section{From classical to quantum solutions}
\label{sec_transition}
The generic form of the  Lagrangian density of the field $\psi$ that
has field terms up to the second order  in its potential energy part is  
\begin{equation}
{\cal L}= {1\over 2} \left (\partial_{\mu} {\psi} \right )\, 
\left (\partial^{\mu} 
{\psi}\right ) - J({\bf x},t)\,\psi - {1\over 2} m^2({\bf x},t)\, \psi^2 \, .
\label{generic}
\end{equation}
To clarify the approach, we consider here a  bosonic 
isoscalar field $\psi\,.$  
This does not limit the consideration and will be generalized
later. 
It is known that  the linear terms $J({\bf x},t)$, often called a
current or force, can always be 
removed from consideration \cite{henley62}.
The time-dependent current  term is responsible for the 
creation of coherent states with a Poissonian  
distribution of particles.  
Intensive studies of
a DCC produced by  linear-type coupling of the pions to the disoriented 
mean field have been performed \cite{camelia97}. Unfortunately,
many problems cannot be easily reduced to this linear approximation. 
As mentioned in  the Introduction,  
for the chiral condensate 
this would require the  Introduction of 
a symmetry breaking  isovector current ${\vec J}({\bf x},t)$, whereas 
introducing a scalar effective mass (a quadratic term in the Lagrangian)
preserves intrinsic symmetries.  

There is one important feature of the problem with a quadratic perturbation in 
the Lagrangian. The equations of 
motion for the field, like Eq. (\ref{fieldequation}), 
are linear for the Lagrangian (\ref{generic}); therefore
they are identical for the classical fields and the corresponding 
operators in the Heisenberg picture. 
Thus, an exact classical solution is related  
to the solution to 
the quantized version of the problem. 
The step from classical to quantum treatment
will be considered next. 

\subsection{Parametric excitation of a harmonic oscillator}
The parametric excitation of 
a quantal 
harmonic oscillator has been extensively considered in the literature 
starting from refs.
\cite{popov69,popov70,baz}. This corresponds to 
our problem in the case  of no spatial
dependence of the effective mass. The presence of
translational symmetry in this case 
leads to the conservation of linear 
momentum. The quantum number of momentum, $k$,
can be used to label the normal modes. Each mode is just a simple
oscillator with the time-dependent frequency. The simple  Fourier
transformation from $x$ to $k$ transforms the Hamiltonian, that corresponds 
to the Lagrangian in Eq. (\ref{generic}), into a sum over modes
(for simplicity we keep one-dimensional notations)
\begin{equation}
H={1\over 2} \int\,dx\, \left ( \dot{\psi}(x,t)^2 + \nabla^2\psi (x,t)
+m^2_{\rm eff} (t) \psi^2 (x,t) \right ) = {1\over 2} \sum_k \left (
\dot{\psi_k}^2+
k^2\,\psi_k^2 + m^2_{\rm eff}(t) \psi_k^2 \, \right ).
\label{hamiltonian}
\end{equation}    
With this reduction we arrive at the problem of the independent development of
a large number of quantum oscillators. With the notation 
$\omega^2_k(t)=m^2_{\rm eff}(t)+k^2\,,$ we have a classical equation of motion 
for each mode
$k$  
\begin{equation}
\ddot{\psi_k}+\omega^2_k (t) \psi_k=0 \, .
\label{classicaloscillator}
\end{equation}
After the quantization,
the  $\psi_k$ and the  corresponding momenta become operators. 
Assuming a  particular normal mode $k$
below we omit this subscript.  

Even for the problem of a simple oscillator, the relation between the classical
and quantum solution is quite subtle; not to mention  that 
classically Eq. (\ref{classicaloscillator}) is analogous to a standard problem 
in quantum mechanics of scattering from a potential $-\omega^2$ 
that is known to have only a limited 
number of cases with exact analytical solutions.  

Classically, the 
$S$-matrix would be defined if we assume that a  solution of 
Eq. (\ref{classicaloscillator}) with the asymptotics in the 
remote past corresponding to the frequency $\omega_{-}\,,$
\begin{equation}
\psi (t)=e^{-i\, \omega_{-} t}\,
\quad {\rm at } \quad t\rightarrow -\infty\, ,
\label{initialstate}
\end{equation}
has 
evolved with
time to a final  general form of the solution with the frequency 
$\omega_{+}\,,$
\begin{equation}
\psi (t)=\,\sqrt{\omega_{-}\over \omega_{+}}
\left (u\, e^{-i\, \omega_{+} t}\,+\, v^{*}\,e^{+i\, \omega_{+} t}
\right )\,
\quad {\rm at } \quad t\rightarrow +\infty\, .
\label{finalstate}
\end{equation}
It is clear that the complex conjugate to Eq. 
(\ref{initialstate}) will develop into a corresponding complex conjugate
version of Eq. (\ref{finalstate}). Furthermore, the unitarity, i.e. 
conservation of probability applied to Eq. (\ref{classicaloscillator}),
results in a restriction posed on $u$ and $v$
\begin{equation}
|u|^2\,-\,|v|^2\,=\,1\,.
\label{uvrestrictions}
\end{equation}
In order to analyze the quantum version of the problem we use the language 
of  secondary quantization, introducing creation and annihilation
operators for a single harmonic oscillator.
The time dependence of a quantized field coordinate $\psi (t)$ in the 
Heisenberg 
representation as
$t\,\rightarrow\, -\infty$ is
\begin{equation}
\psi(t)\,=\,{1\over \sqrt{2\omega_{-}}}\left ( a\,e^{-i\,\omega_{-}t}\,+\,
a^{\dagger}\,e^{i\,\omega_{-}t}\right )\, ,
\label{instate}
\end{equation}
where the operators $a$ and $a^{\dagger}$ do not depend on time and
define the ``in'' state of the system.
We will define the ``out'' state using operators  $b$ and $b^{\dagger}$
as
\begin{equation}
\psi(t)\,=\,{1\over \sqrt{2\omega_{+}}}\left ( b\,e^{-i\,\omega_{+}t}\,
+\,b^{\dagger}\,e^{i\,\omega_{+}t}\right )\, \quad {\rm at} 
\quad t\,\rightarrow\, +\infty\,.
\label{outstate}
\end{equation}
Being supported by the fact that the field $\psi (t)$, even in its quantum 
form, 
satisfies  Eq. (\ref{classicaloscillator}), and with our assumptions
about the classical solution  (\ref{initialstate}, \ref{finalstate})
we can  relate operators $a$ and $a^{\dagger}$ with 
$b$ and $b^{\dagger}\, ,$
\begin{equation}
b\,=\,\left ( u\, a + v\, a^{\dagger}
\right )\, ,
\quad b^{\dagger}\,
=\, \left ( u^{*}\, a^{\dagger} 
+ v^{*}\, a \right ) \, ,
\label{bogoliubov}
\end{equation}
with the same parameters $u$ and $v$ as in the classical solution.
As seen  (\ref{bogoliubov}) the ``in'' and ``out'' states are related by
a Bogoliubov transformation. The condition that the commutation relation 
$$[b,b^{\dagger}]\,=\,( |u|^2-|v|^2)\,
[a,a^{\dagger}]\,=\,1
$$
is preserved coincides with Eq. (\ref{uvrestrictions}) .

It is possible to 
build an $S$-matrix corresponding to the transformation in 
Eq. (\ref{bogoliubov}),
however the solution is very technical, and of limited use. We can use
Eq. (\ref{bogoliubov}) directly to answer simple but relevant questions.
The probability amplitude to have $n$ final quanta if there was vacuum in the 
initial state is given by
$$C_n\,=\, _{a}\langle n|S|0\rangle_{a}\,=\,_{a}\langle n|0\rangle_{b}\, ,$$
where the $S$-matrix is a transition matrix from the 
``in'' state to the ``out'' state.
Then $$|0\rangle _{b}\,=\, \sum_{n=0}^{\infty}\, C_n |n\rangle _a\, ,$$
and recursion  relation for the coefficients $C_n$
\begin{equation}
C_{n+2}=-{v\over u}\,\sqrt{n+1\over n+2}\, C_n\,  
\label{recursion}
\end{equation}
may be found from the definition
of the vacuum and Eq. (\ref{bogoliubov}):
$$b|0\rangle_{b}\,=\,\sum_{n=0}^{\infty}   
C_n\,\left (u\, \sqrt{n} \,|n-1\rangle_{a} + v\, \sqrt{n+1} \,|n+1\rangle_{a}
\right )\,=\,0\,.$$
Finally, normalized to unity, the transition probability from a vacuum to a 
state with $2n$ bosons is expressed by
\begin{equation}
|C_{2n}|^2\,\equiv \,P(2n)\,=
\,{(2n)!\over 2^{2n}\, (n!)^2}\, \rho^n\,\sqrt{1-\rho}\, ,
\label{transition}
\end{equation}
where the parameter $\rho=|v/u|^2$ represents 
the reflection probability  for  Eq. 
(\ref{classicaloscillator}) if it is viewed as a Schr\"odinger equation 
for scattering at zero energy off a potential $-\omega^2\,.$
Conservation of probability, Eq. (\ref{uvrestrictions}), limits the values
of $\rho$ to the region $\rho\in [0,1)\,.$
The distribution of particles produced from vacuum 
Eq. (\ref{transition}) peaks at zero
and the average number of particles produced is
\begin{equation}
\overline{n}\,=\,\sum_{n=0}^{\infty}\, 2n\, P(2n)\,=\, 
{\rho\over 1-\rho}\, .
\label{number1d}
\end{equation}
It is seen from the above expression that the number of particles diverges
as $\rho$ approaches one. This will be a region of interest
in our later studies of chiral condensates. 

A rigorous and detailed discussion of all the features in the transition
from classical solutions of Eq. (\ref{classicaloscillator}) to a problem of 
a quantum oscillator with a time-dependent frequency 
can be found in \cite{popov69,popov70,baz}.
The corresponding  Schr\"odinger equation for the wave function 
of the quantum oscillator is ($\hbar =1$) 
\begin{equation}
i\,{\partial \Phi (x,t) \over \partial t}\, =\, 
- \, {1\over 2}\,{\partial^2 \Phi (x,t) \over 
\partial x^2}+  {1\over 2}\, \omega^2(t)\, x^2\, 
\Phi (x,t) \, ,
\label{quantumoscillator}
\end{equation}
where coordinate $x$ originates from  the field ``coordinate'' $\psi$, Eq. 
(\ref{classicaloscillator}). 
Let some function $\psi(t)$ be a solution to classical Eq. 
(\ref{classicaloscillator}) with  
$\omega(t=-\infty)=\omega_{-} $ and  $\omega(t=+\infty)=\omega_{+}$.
This function  
$\psi(t)$ can be written in terms of the real part $r(t)$ and 
phase $\gamma(t)$ so that  $\psi (t)= \, r(t)\, e^{i\gamma(t)}\, .$ With a 
direct substitution it can be shown that  
\begin{equation}
\Phi(x,t)={1\over\sqrt{r(t)}}\, \exp{\left ( i{\dot{r}x^2\over 2 r}\right )}
\, \chi (y, \tau)
\label{quantumsolution}
\end{equation}
is a solution to the quantum  equation 
(\ref{quantumoscillator}). Here a rescaled length $y$ and time $\tau$ are
introduced as $y=x/r(t)\, ,$ $\tau=\gamma (t)/\omega _{-} $. The function 
$\chi(x,t)$ is a standard solution of the Schr\"odinger equation for the
oscillator with a constant frequency $\omega_{-}\,,$
\begin{equation}
i\,{\partial \chi (y,\tau) \over \partial \tau}\, =\, 
- \, {1\over 2}\,{\partial^2 \chi (y,\tau) \over 
\partial y^2}+ {1\over 2}\, \omega_{-}\, y^2\, 
\chi (y,\tau)\, .
\end{equation}
The initial conditions for the solution $\chi (y,\tau)$ 
should be set at $t\rightarrow -\infty$ by the known conditions
on $\Phi(x,t)\,.$
Equation (\ref{transition}) can be generalized 
utilizing this more general technique.
It can be shown
that the  probability  for a transition
from the state with $m$ quanta into the final state of $n$ quanta is given by
\begin{equation}
\,{\min(m,n)!\over \max(m,n)!} \sqrt{1-\rho} 
\left | P_{(m+n)/2}^{|m-n|/2} (\sqrt{1-\rho})\right |^2 \, ,
\label{generaltransition}
\end{equation} 
where $P_{\nu}^{\mu}$ are the associated Legendre polynomials and $n$ and $m$ have
to be of the same parity.
The established bridge between classical and quantum solutions is a 
remarkable achievement, but unfortunately analysis of
Eq. (\ref{quantumsolution}) 
does not present a pleasant task.

\subsection{Infinite number of mixed modes}
\subsubsection{General canonical transformation}
The situation becomes more complicated if we return to the general
field equation given by the Lagrangian density of Eq. (\ref{generic}), 
and assume that the modes cannot be
separated. We keep assuming $J(x, t)=0$ which is relevant to our particular
problem but in general this factor can be included back in the discussion
without much complication. The problem of the system of coupled oscillators 
has been discussed
in  great detail in \cite{mollow67}. The transformation  
analogous
to Eq. (\ref{bogoliubov}) now takes an $N$-dimensional symplectic form,
where $N$ is the number of coupled modes \cite{blaizot86},
\begin{equation}
b_k\,=\,\sum_{k^{\prime}=1}^N \, \left ( u_{k\,k^{\prime}}\, a_{k^{\prime}} + 
v_{k\,k^{\prime}}\, a_{k^{\prime}}^{\dagger} \right )\, ,
\quad b_{k}^{\dagger}\,
=\, \sum_{k^{\prime}=1}^N \, \left 
(u_{k\,k^{\prime}}^{*}\, a_{k^{\prime}}^{\dagger} 
+ v_{k\,k^{\prime}}^{*}\, a_{k^{\prime}} \right ) \, ,
\label{generalbogoliubov}
\end{equation}
It is convenient to regard the operators $a_k$ as components of 
an $N$-dimensional 
vector, and the 
numbers $u_{k\,k^{\prime}}$ and $v_{k\,k^{\prime}}$ as 
$N\times N$ matrices.   
Eq.  
(\ref{generalbogoliubov}) in matrix notation takes the form
\begin{equation}
b\,= S^{-1}\,a\, S\,=\,u\,a\,+\,v\,a^{\dagger}\, ,\quad
b^{\dagger}\,=\,S^{-1}\,a^{\dagger}\, S\,=\,u^{*}\,a^{\dagger}\,+
\,v^{*}\,a\,.
\label{matrixu1}
\end{equation}     
Using the properties of the matrices $u$ and $v$ shown 
below, it is straightforward to check that the inverse transformation is 
given by
\begin{equation}
a\,= S\,b\, S^{-1}\,=\,u^{\dagger}\,b\,-\,v^{T}\,b^{\dagger}\, ,\quad
a^{\dagger}\,=\,S\,b^{\dagger}\, S^{-1}\,=\,u^{T}\,b^{\dagger}\,-
\,v^{\dagger}\,b\,,
\label{matrixu2}
\end{equation}
where $u^{*}$, $u^{\dagger}$ and $u^{T}$ have their usual meanings of
complex conjugate, hermitian conjugate and transpose matrices, respectively.
Further, we will also use an inversion denoted as $u^{-1}\,.$
Similarly to the one-mode example of 
Eq. (\ref{uvrestrictions}), matrices 
$u$ and $v$ are subject to conditions that arise from the fact that
the commutation relations have to be preserved. It follows from 
Eq. (\ref{matrixu1}) that
\begin{equation}
u\, u^{\dagger}\,-\, v\, v^{\dagger}\,=\,1\, , \quad u\, v^{T}\,-\,v\, u^{T}=0
\, ,
\label{generalcommutaions}
\end{equation}
and from the inverse transformation, Eq. (\ref{matrixu2})
\begin{equation}
u^{\dagger}\, u\,-\, v^{T}\, v^{*}\,=\,1\, , \quad u^{\dagger}\, v\,
-\,v^{T}\, u^{*}=0 \, .
\label{generalcommutaions1}
\end{equation}
\subsubsection{Transitions between coherent states}
In the case of many mixed modes it is still possible to obtain recursion
relations similar to  the single-mode situation of Eq. (\ref{recursion}) 
but they become  difficult to 
analyze. Next we will discuss the approach found in \cite{mollow67}
with a different technique that uses 
the transitions between coherent  states.
We define a single-mode coherent state $|\alpha \rangle$ in the usual way as
\begin{equation}
|\alpha\rangle\,=\,e^{-|\alpha |^2/2}\,\sum_{n=0}^{\infty}\,{\alpha^n\over
\sqrt{ n!}}
|n\rangle\, .
\label{defenition1}
\end{equation}
In mathematics sums as in Eq. (\ref{defenition1}) 
are often called generating functions.
It is convenient to introduce  states with a different normalization
\cite{glauber63} as
\begin{equation}
||\alpha\rangle\,=\,e^{|\alpha|^2/2}\,|\alpha\rangle\,=\,e^{\alpha\,
a^{\dagger}}\,|0\rangle \, .
\label{defenition2}
\end{equation}
When applied to the states (\ref{defenition2}), the creation operation is 
equivalent to the derivative,
\begin{equation}
a^{\dagger}\,||\alpha\rangle\,=\, {\partial\over\partial \alpha}\, ||\alpha
\rangle \, .
\label{property}
\end{equation}

We continue to
use the notation $|\bbox{\alpha}\rangle$ for a multidimensional coherent 
state, which should be interpreted as a product of the single-mode states.
$\bbox{\alpha}$ is a vector and the derivative ${\partial/
\partial \bbox{\alpha}}$ is understood as a gradient in $N$-dimensional space 
of modes. 
We are interested in finding the matrix elements of the evolution matrix
$S$ that implements the unitary transformation between initial and final states
of the system. The evolution of the 
initial state is quite complicated and unless
$v=0$ the coherent state does not stay coherent
as in general it evolves into a so-called
``squeezed'' state \cite{picinbono70,zhang90}. 
Nevertheless it is possible to obtain
an analytic expression for the matrix elements of the evolution operator $S$
between coherent states. 
To proceed in this direction we will consider the action of the
matrix $S$ on the creation and annihilation operators given by Eqs. 
(\ref{matrixu1}, \ref{matrixu2}), that actually 
serve here as the definition of the evolution matrix. 
By acting on the complex conjugate form of the first equation in Eq. 
(\ref{matrixu1}) with $\langle \bbox{\beta} \,|| S$ 
from the left and with $ ||\,\bbox{\alpha} \, \rangle$ from the right, and 
utilizing Eq. (\ref{property}) we arrive at the following differential 
equation
\begin{equation}
\left ( \bbox{\beta}^{*}\,-\,u^{*}\, {\partial \over \partial \bbox{\alpha}}\, -\,
v^{*}\, \bbox{\alpha}\, \right ) \langle \bbox{\beta} ||S||\bbox{\alpha}\rangle \,=\,0\, .
\label{dif1}
\end{equation}
In a similar manner from the first equation of Eq. (\ref{matrixu2})
we obtain
\begin{equation}
\left ( \bbox{\alpha}\,-\,u^{\dagger}\, {\partial \over \partial \bbox{\beta}^{*}}\, +\,
v^{T}\, \bbox{\beta}^{*}\, \right ) \langle \bbox{\beta} ||S||\bbox{\alpha}\rangle \,=\,0\, .
\label{dif2}
\end{equation}
The solution to the differential equations (\ref{dif1},\ref{dif2})
determines the transition amplitude up to a normalization constant $C(u,v)$:
\begin{equation}
\langle \bbox{\beta} ||S||\bbox{\alpha}\rangle \,=\,C(u,v)\, \exp \left (\bbox{\beta}^{\dagger}\,
(u^{\dagger})^{-1}\,\bbox{\alpha}\,+\,{1\over 2}\, \bbox{\beta}^{\dagger}
\,v\,(u^{*})^{-1}\,
\bbox{\beta}^{*}
\,-\,{1\over 2}\,\bbox{\alpha}^{T}\, (u^{*})^{-1}\, v^{*}\, \bbox{\alpha}
\,\right )\, .
\label{amplitude}
\end{equation}
This solution can be checked directly by substitution into 
Eqs. (\ref{dif1},\ref{dif2}) and utilizing the observation which follows from
Eqs. (\ref{generalcommutaions}, \ref{generalcommutaions1}) that matrices
$v\,(u^{*})^{-1}$ and  $(u^{*})^{-1}\, v^{*}$ are both symmetric.
The normalization constant may be obtained using the completeness of coherent 
states \cite{blaizot86}, 
\begin{equation}
C(u,v)=\left( \det(u u^{\dagger}) \right)^{-1/4}\, .
\label{normalization}
\end{equation} 
The evolution matrix given in the form of Eq. (\ref{amplitude}) can be 
transformed via a Taylor expansion 
in the particle number basis using the definitions in
Eqs. (\ref{defenition1}, \ref{defenition2})
\begin{equation}
\langle \bbox{\beta} ||S||\bbox{\alpha} \rangle\, =\, \sum_{\{n_k\}\,\{n^{\prime}
_{l}\}}\,
\langle {n_k} |S| {n^{\prime}_{l}} 
\rangle\, \prod_{k}^N\, {{\beta_k^{*}}^{n_k}
\over \sqrt{{n_k}!}}\,\prod_{l}^N\, {\alpha_l^{n^{\prime}_{l}}
\over \sqrt{n^{\prime}_{l}}!} \,.
\end{equation} 
In general it is quite complicated to give a finite expression 
for the coefficients of 
the Taylor series arising from the multi-variable Gaussian.
Expansions of two-variable Gaussians are known to be of the
form of Legendre polynomials which give rise to  Eq.
(\ref{generaltransition}).  
Nevertheless, the algorithm for the expansion is straightforward, first the
exponent should be expanded in terms of its argument and then
each term can be 
expanded into a final sum with a generalized binomial expansion.

By considering a transition from the vacuum state we set all terms with 
$\bbox{\alpha}$ to zero in Eq. (\ref{amplitude}). In this case everything is 
completely determined by the matrix $v\,(u^{*})^{-1}\,.$ For the question
of multiplicity distributions in the one-mode case, we  note that 
the relative phase between $u$ and $v$ was of no importance, a single 
parameter $\rho$ determined everything. We will further see that a 
similar picture
holds in the general case, and only one matrix $v v^{\dagger}$ is needed to 
find  the particle distributions. With the help of Eqs.
(\ref{generalcommutaions}, \ref{generalcommutaions1}) it can be seen that
the matrices $u u^\dagger$ and $v v^\dagger$ can be diagonalized simultaneously
and the eigenvalues of $v v^\dagger(u u^\dagger)^{(-1)}=
v\,(u^{*})^{-1}(v\,(u^{*})^{-1})^{\dagger}$  form a set of 
parameters $\rho$ different for each mode.

\subsubsection{Multiplicity distributions}  
Despite the complicated form of a general expression  (\ref{amplitude})
one may calculate the moments of multiplicity distributions in a 
straightforward manner. 
Let the  initial state $|i\rangle$ be characterized
by a diagonal density matrix $\hat{n}^{\circ}$, 
a set of numbers of quanta $\{n^{\circ}_k\}$ in each mode $k$ on the diagonal.
The average number of quanta in a final state $|f\rangle$ 
is determined by 
$$\overline{n}_k\,=\,\langle f |a^{\dagger}_k\, 
a_k|f \rangle\,=\,\langle i|b^{\dagger}_k
\,b_k|i\rangle \, .$$
With the above assumptions and Eqs. (\ref{matrixu1}, \ref{matrixu2}),
the 
density matrix $\hat{n}$ of a final state is
\begin{equation}
\hat{n}\,=\,u \hat{n^{\circ}} u^{\dagger} + v \hat{n^{\circ}} v^{\dagger}+ 
v v^{\dagger}\,.
\label{numbergeneral}
\end{equation} 
Throughout the rest of the work 
we will concentrate on the situation of a particular interest
when particles are created from the vacuum and  
the first two terms in Eq. 
(\ref{numbergeneral}) are identically zero. The average total number of 
particles in this case may be expressed in a simple matrix form
\begin{equation}
\overline{n}_{\rm total}\,=\,\sum_k\,n_k\,=\,{\rm Tr} (v^{\dagger} v) \, ,
\label{number}  
\end{equation}
which is consistent with Eq. (\ref{number1d}).
Higher moments of the particle distributions can be 
calculated in the same manner. 
Unfortunately, the calculation of an arbitrary moment requires
path integration techniques while using Wick's theorem in the
normal ordering of operators.  Low order moments
can be directly  calculated, for example
\begin{equation}
\overline{n^2}_{\rm total}\,=\, \overline{n}_{\rm total}^2\,+
\,2\overline{n}_{\rm total}\,+
\,2\,{\rm Tr}\left ((v v^{\dagger}\,)^2\right )\,,
\label{moments_a}
\end{equation}
\begin{equation}
\overline{n^3}_{\rm total}\,=
\overline{n}_{\rm total}^3+
6\overline{n}_{\rm total}^2+
4\overline{n}_{\rm total}
+6\overline{n}_{\rm total}\,{\rm Tr}\left ((v v^{\dagger}\,)^2\right )\,
+12 {\rm Tr}\left ((v v^{\dagger}\,)^2\right )
+8 {\rm Tr}\left ((v v^{\dagger}\,)^3\right )\,.
\label{moments}
\end{equation}
Equation (\ref{moments_a}) can be identified as a super-Poissonian distribution
of particle pairs.
It follows from the above expressions  that
particle production is determined by the matrix $v v^\dagger$ which is 
related in a simple way to $u u^\dagger\,,$ Eq. (\ref{uvrestrictions}).
As was mentioned before,  
the hermitian  matrix $v v^{\dagger}$ can always be diagonalized with 
diagonal elements being average numbers of particles $\overline{n_q}$ 
in each eigenmode $q$. This diagonalization allows us to view particle 
production as a production from independent modes. 
A connection can be established  to the one-oscillator case, previously 
considered, by defining a
set of parameters $\rho$ as $\rho_q=\overline{n_q}/(1+\overline{n_q})\,,$
that are the eigenvalues of $v v^{\dagger} u u^{\dagger}\,.$ 
Considering the number of particles produced in each eigenmode
we can restrict ourselves by the modes with $\rho_q \rightarrow 1\,,$ 
that dominate the particle production.
We will later refer to these modes as condensate modes.
In the following
section, it will be shown that 
the distinct physical feature of these modes is that they produce an 
exponentially large number of particles.

Next we consider a number of condensate modes with equal parameters
$\rho_q$. For one mode the  answer is in  Eq. ({\ref{transition}). 
For the case with
several condensate modes  one  needs to know the 
distribution for the total sum of the particles, which is given by the
convolution of the corresponding probabilities.
As for the Fourier transformation, 
a  convolution of several 
distributions results in a product of their generating functions.
The generating function for the distribution  (\ref{transition}) is
$$ \tilde{P}(y)\,=\,\sum_{n=0}^{\infty}\,P(2n)\, y^n \,=\,
{\sqrt{1-\rho}\over\sqrt{1-\rho y}}\, . $$
We obtain total probabilities
of particle production for any combination of species or any number of modes 
convoluted together, as a Fourier expansion of a
product of corresponding generating functions.
Suppose we are looking for a distribution $P_l(2n)$that gives a probability 
to observe $2n$  particles appearing in $l$ single-mode
distributions  (\ref{transition}).
The Taylor expansion of $(1-\rho y)^{-l/2}$ gives 
\begin{equation}
P_l(2n)\,=\,{(l+2(n-1))!!\over 2^n\,n!\, (l-2)!!}\,\rho^n\,(1-\rho)^{l/2} \,,
\label{convolution}
\end{equation}
where the double factorial should be understood appropriately for odd and even 
$l\,.$

According to the central limit theorem, only the distributions corresponding to
a small number of condensate modes have a shape that is very different
from a Gaussian, see Figure \ref{oneD} below. 
If parameters $\{\rho_q\}$ for the condensate modes are very different then
only the important modes that produce many particles can be isolated
reducing the problem back to the case of several almost degenerate modes.
The method of generating functions is also convenient in discussions
of distributions over species. For example,  
the distribution for the  total number of pions, regardless of the 
isospin projection, 
can be obtained as a distribution of the sum of three species, 
via convolution. 
Alternatively, one can 
think of the total condensate modes for pions as 
just a sum of numbers of condensate 
modes for all species.    

As an example we apply Eq. (\ref{convolution}) for one mode and three pion 
species. The  probability of having $2n$ neutral pions is
\begin{equation}
P^{0}(2n)\,=\,P_1(2n)\,=\,{(2n)!\over 2^{2n}\, (n!)^2}\, 
(1-\rho)^{1/2}\,\rho^n\, ,
\label{pineutral}
\end{equation}
the probability for observing $2n$ charged pions is
\begin{equation}
P^{+}(n)\,=\,P^{-}(n)\,=\,P^{\rm charged}(2n)\,\,=\,P_2(2n)\,
=\,(1-\rho)\rho^n\, ,
\label{picharged}
\end{equation}
and finally the probability of having total $2n$ pions is
\begin{equation}
P^{\rm tot}(2n)\,=P_3(2n)\,=
\,{(2n+1)! \over 2^{2n}\, (n!)^2}\,(1-\rho)^{3/2}\,\rho^n \,.
\label{pitotal}
\end{equation}
We note the convolution of three distributions in Eq. (\ref{pineutral}) that
give rise to Eq. (\ref{pitotal}) produces a peaked curve with maximum at
particle number 
$$n_{\rm max}\,=\,{3\rho-2\over 1-\rho}\, .$$ 
The probability distributions (\ref{pineutral},
\ref{picharged}, \ref{pitotal}) are shown in Fig. \ref{oneD} with solid
line, long dashed line and short dashed line, respectively. 
The figure displays 
a critical situation when parameter $\rho$ is 0.999. This corresponds closely
to the first mode in the example of a square perturbation considered in the 
next section.  
\begin{figure}
\begin{center}
\epsfxsize=9.0cm \epsfbox{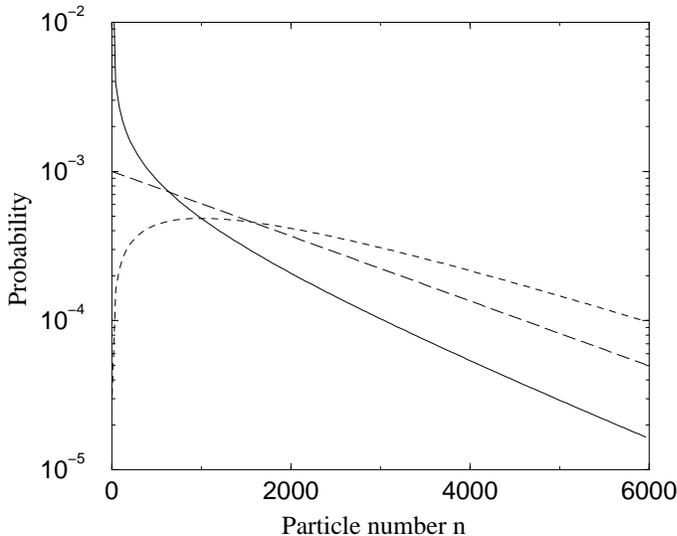}
\end{center}
\caption{
The particle number distribution 
for neutral pions (solid line), charged pions (long dashes), 
and for all pions (short dashed line).
All curves are normalized to unity; note that the  
total number of pions and number of neutral pions both are always even, while 
any integer number of charged pions appearing in pairs 
is allowed. The parameter $\rho=0.999\,.$
\label{oneD}}
\end{figure}

Now we can return to  original questions: 
What is the probability to observe a certain 
fraction of neutral  
pions given a total number of pions detected? Is it
possible to use this distribution for the detection of chiral condensate? 
In the one-mode approximation,
given a total number of pions produced $2n_t\,,$ the normalized probability
of observing $2n$ neutral particles is
\begin{equation}
{\cal P}
\left (f={n\over n_t}\right )\,=\,{ P^{0}(2n)\,P^{\rm charged}(2n_t-2n)\over
P^{\rm tot}(2n_t)}\,=\,{(2n)!\over (2n_t)!}\, \left ({n_t!\over n!} \right )^2
\, 2^{2n_t-n} \,.
\label{probability1}
\end{equation}
Finally, if both $n$ and $n_t$ are large, which is a  good approximation
in almost all regions, Stirling's formula may be used, and one finds
\begin{equation}
{\cal P}(f)\,=\, {1\over (2n_t+1)\sqrt{f}} \, ,
\label{smode}
\end{equation}
which coincides with Eq. (\ref{probability}) if the normalization over $n_t+1$
discrete points between zero and one 
is switched to the integral over $f\in [0,1]$.  

Unfortunately, this result
does not hold for all situations when several  
modes are participating together. 
The single-mode result (\ref{smode}) may be invalid
when 
the largest eigenvalue of the matrix $v v^\dagger$ is exactly or 
almost  degenerate. Physically, this may be due to  some
symmetry for example.  
One- and two-mode cases have their 
probability peaked at zero, this is no longer 
true for the convolution of three or more modes, Fig. \ref{oneD}.
Eq. (\ref{convolution}) can be applied to give a result
for any number of modes $j$ participating in the condensate, assuming they all 
have equal strength $\rho\,.$
The distribution of neutral pions is given by 
\begin{equation}
{\cal P}_j
\left (f={n\over n_t}\right )\,=\,{ P_j(2n)\,P_{2j}(2n_t-2n)\over
P_{3j}(2n_t)}\, .
\label{exactf}
\end{equation}
With a Stirling's formula and the assumption  $n\gg j\,,$ 
Eq. (\ref{exactf})
can be simplified to 
\begin{equation}
{\cal P}_j(f)\,=\,{(3j/2-1)!\over (j/2-1)!\,(j-1)!}\,f^{j/2-1}\,(1-f)^{j-1}\, . 
\label{approxf}
\end{equation} 
For comparison, the probability distributions for one, two, and three
modes in the condensate are shown in Fig. \ref{poff}.
\begin{figure}
\begin{center}
\epsfxsize=9.0cm \epsfbox{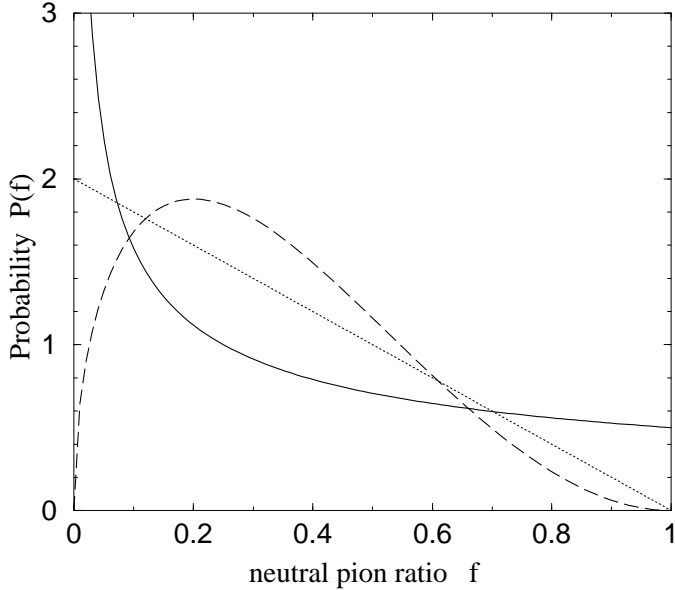}
\end{center}
\caption{The  probability  ${\cal P}(f)$ that a given neutral pion
fraction $f$ is observed. The three curves display cases of one 
condensate mode (solid line), two energy-degenerate modes (dotted line),
and three modes (dashed line). 
\label{poff}}
\end{figure}
The distributions of Eq. (\ref{approxf}) have maxima, average and widths 
as follows
$$f_{\rm max}\,=\,{
j-3\over 3j-4}\,,\,\,\,
\overline{f}=1/3\,,\,\,\, 
\overline{f^2}-\overline{f}^2\,=\,{4\over 27j+18}\, ,$$
where the first equation can be applied with the restriction  $j>2\,.$ 
It follows from the central limit theorem that
for a large number of modes one should expect a Gaussian distribution 
that tends to a $\delta$-function. Thus, as the number of modes in the 
condensate
grows, there is a fast transition from a $1/\sqrt{f}$ type behavior to
a sharp  peak at $f=1/3\,.$

\subsubsection{Dynamics of matrices $u$ and $v$}
As some final remarks about the link between classical and quantum
solutions to the wave equation arising from the Lagrangian in 
Eq. (\ref{generic}),
we would like to discuss the technical question of constructing matrices
$u$ and $v\,.$
Unfortunately,  Eq. (\ref{fieldequation}) has no general analytic 
solution even
classically.
The Green's function formalism reduces the problem
to a Fredholm type integral equation. The exact solution can be obtained only
in special cases, for  a separable kernel \cite{henley62}, or
in the sudden perturbation limit to be considered in the next section.
However, numerical studies of Eq. (\ref{fieldequation}) seem
to offer a great chance of success. 

It would be good to obtain equations for $u$ and $v$ that are still
exact but written in the form convenient for numerical
work. 
Let us quantize the field at every intermediate stage with bare 
particles \cite{henley62} so that
\begin{equation}
\psi(x,t)=\sum_k\, {1\over \sqrt{2\, \omega_k\, L}}
\left ( b_k(t)\, e^{i\,k\,x\,-\,i\omega_k\,t}\,+
\,b_k^{\dagger}(t)\, e^{-i\,k\,x\,+\,i\omega_k\,t} \right )\, ,
\label{bare}
\end{equation}  
where $b_k(t)$ is now a time-dependent annihilation operator. 
Eq. (\ref{bare}) is written in an interaction representation, 
the explicit time dependence of the free field is given by 
the exponents, whereas creation and annihilation operators absorb
the remaining nontrivial time dependence. 
In Eq. (\ref{bare}) the variable $\omega_k\,=\left (\,k^2\,+\,
m^2_{\it eff}(t\rightarrow -\infty)\right )^{1/2}$ 
is the initial time-independent frequency, thus there is no problem if 
at any point in time the effective mass goes through zero.
As time goes to infinity $\psi(x,t)$ becomes the 
final ``out'' state, with operators  $b$ and $b^{\dagger}$ defined as before.
For  
one-dimensional case  $L$ 
is the quantization length. 
For further simplicity we to denote 
\begin{equation}
m^2_{\it eff}(x,t)\,=\,
m^2_{\it eff}(t\rightarrow -\infty)\,+\, \Pi(x,t)\, ,
\label{defPi}
\end{equation}
this allows the
separation of the interaction Hamiltonian.
The perturbed Hamiltonian from Eq. (\ref{hamiltonian}) expressed in terms 
of $b(t)$ and $b^{\dagger}(t)$ is
\begin{equation}
H^{\it int}(t)=\sum_{k\,k^{\prime}}\,\left ( \Omega_{k\,k^{\prime}}(t)\left ( 
b_k^{\dagger} (t)\,
b_{k^{\prime}}(t)+{\delta_{k\,k^{\prime}}\over 2}\right )\,+\,{1\over 2}\,\Lambda_{k\,k^{\prime}}(t)\,b_k^{\dagger} (t)  
b_{k^{\prime}}^{\dagger}(t)+{1\over 2}\,\Lambda_{k\,k^{\prime}}^{*}(t)
\,b_k (t)b_{k^{\prime}}(t)\,  \right )\,,
\end{equation}
where the  matrices $\Omega$ and $\Lambda$ are determined as follows:
\begin{equation}
\Omega_{k\,k^{\prime}}(t)\,=\,
{1
\over 2 L \sqrt{\omega_k \omega_{k^{\prime}}}}\, \Re \left (
\Pi(k-k^{\prime},\, t)e^{i(\omega_{k^{\prime}}-\omega_k)t} \right )\, ,\,\, 
\Lambda_{k\,k^{\prime}}(t)\,=\,{\Pi(-k^{\prime}-k,\,t)\over 2 L\sqrt{\omega_k 
\omega_{k^{\prime}}}}\,e^{i(\omega_{k^{\prime}}+\omega_k)t}\,.
\end{equation}
In the above expression $\Pi(k,t)$ is a Fourier image of $\Pi(x,t)\,,$
determined as
$$\Pi(k,t)\,=\,\Pi^{*}(-k,t)\,=\,
\int\,\Pi(x,t)\, e^{i\,k\, x} \, dx \,,$$
and $\Re$ denotes a real part of the expression.
Utilizing the 
Hamiltonian equation of motion in the interaction picture
$$i\,{d\over dt}\, b(t)\,=\, [b(t),H^{\it int.}(t)]\,=\,
\Omega \, b(t)\,+\, \Lambda\, b^{\dagger}(t) $$
it is possible to show
that if $b(t)$ is defined through the initial operators $a$ and $a^{\dagger}$
as 
\begin{equation}
b(t)\,=\,u(t)\,a\,+\,v(t)\,a^{\dagger}\, ,
\label{matrixtime}
\end{equation}
where $u$ and $v$ must satisfy the matrix equations
\begin{eqnarray}
\label{uvequation1}
i\,{d\over dt}\, u(t)\,=\,\Omega(t)\,u(t)\,+\,\Lambda (t) v^{*}(t)\,,\\
i\,{d\over dt}\, v(t)\,=\,\Omega(t)\,v(t)\,+\,\Lambda (t) u^{*}(t)\,.
\label{uvequation2}
\end{eqnarray}
At infinitely large times when the perturbation $\Pi$
goes to zero, the right hand 
sides of Eqs. (\ref{uvequation1}, \ref{uvequation2})
vanish and $u$ and $v$ become time-independent. In order to 
obtain matrices $u$ and $v$ for Eq. (\ref{matrixu1}) 
one has to solve the above equations with 
initial conditions $u=1$ and $v=0\, .$

\section{Application to chiral condensate}
\label{sec_application}

\subsection{Separable modes, space-independent effective mass}
To illustrate the  machinery developed in the 
previous section we start with simple  cases when the  
classical solution is known analytically.  
The first  example 
is the space-independent field
$\Pi ({\bf x},t)\equiv\Pi(t)\,,$ see Eq. (\ref{defPi}), 
i.e. the
situation when the perturbation is uniform in a  box to which 
the entire pion field is confined. 
The wave vector $\bf{k}$ is a good quantum number. 
Particles get produced independently 
in each mode
labeled with $\bf{k}$, and production is determined by
the classical reflection probability $\rho_{\bf k}\,=\,|v_{\bf k}|^2/
|u_{\bf k}|^2\,,$ Eq. (\ref{transition}). The distribution of particles for
a single mode, Eq. (\ref{transition}), is a decaying 
function that has a maximum at zero. Its behavior
can be approximated with Stirling's formula as
$$P(2n)\,\sim\, {\rho^n\over \sqrt{n}}\, , $$
where   $P(2n)$ denotes the 
probability of creating $2n$ particles from the vacuum. 
The  distribution in Eq. (\ref{transition}) 
for a value of $\rho=0.999$ 
is shown in Fig. \ref{oneD} as a solid line; 
some additional discussion is given below. 
The average number of
particles produced in Eq. (\ref{number1d}) is generally quite small unless 
we are in the condensate region when $\rho\rightarrow 1\,.$
With the assumption of independent modes Eq. (\ref{eqmotion}) reduces
to 
\begin{equation}
-{d^2 \pi_{\tau}/ d t^2} - \Pi(t)\pi_{\tau}=(m_{\pi}^2+k^2)\pi_{\tau}\,.
\label{shrodinger}
\end{equation}
Eq. (\ref{shrodinger}) is written in  a form of the 
Schr\"odinger equation for scattering from a potential
barrier of height $-\Pi$ at ``energy'' $m_{\pi}^2+k^2\,.$ 
The parameter $\rho$  we are looking for, which
links classical and quantum pictures, is the reflection coefficient 
for this scattering process. 

Let the effective pion mass  
change and then return back to normal in a step function manner.
This corresponds to the scattering off a rectangular potential barrier. 
The situation when the tunneling is involved, is relevant to the case of 
a low momentum mode being amplified as the index of reflection is 
rapidly increasing.
We assume that the perturbation $\Pi (t)$ has a non-zero 
value $\Pi$ only for the time interval $t\in [0,T]\,.$
The reflection probability for this scattering potential 
is 
$$\rho_{\bf k}\,=\,{|v_{\bf k}|^2\over 1+|v_{\bf k}|^2}\,, $$
and the average number $\overline{n}_{\bf k} $ of particles created 
in the mode ${\bf k}\,$ is given by
\begin{equation} 
\overline{n}_{\bf k}\,=\,|v_{\bf k}|^2\,=
\,{\Pi^2\over 4\,(m_{\pi}^2\,+\,{\bf k}^2)\,|m_{\pi}^2\,+
\,{\bf k}^2\,+\,\Pi |}\,\left |\,\sin\left (T \sqrt{m_{\pi}^2\,+
\,{\bf k}^2\,+\,\Pi } \right ) \right |^2\, , 
\end{equation}       

Another form of the perturbation $\Pi (t) $ for which  
an exact analytical solution exists is the Eckart potential \cite{eckart30}
$$\Pi (t)\,=\, {\Pi \over \cosh^2(t/T)}\, ,$$
where $\Pi=\Pi(0)$ is the minimum value of  $\Pi(t)$ and $T$ is 
the time scale of perturbation. 
The resulting form of the reflection probability is 
$$\rho_{\bf k}\,=\, \left | { 1+\cos \left (\pi\, 
\sqrt{4\,\Pi\,T^2\,+\,1} \right )\over
\cosh \left (2\,\pi\,T\,\sqrt{m_{\pi}^2\,+\,{\bf k}^2}\right )\,+\,
\cos \left (\pi\, \sqrt{4\,\Pi\,T^2\,+\,1}\right ) } \right | $$
Both forms of the perturbation $\Pi(t)$ with parameters that we use below for 
our numerical estimates are plotted in Fig. \ref{potentials}, right
side shows the Eckart potential, rectangular barrier is on the left.
\begin{figure}
\begin{center}
\epsfxsize=9.0cm \epsfbox{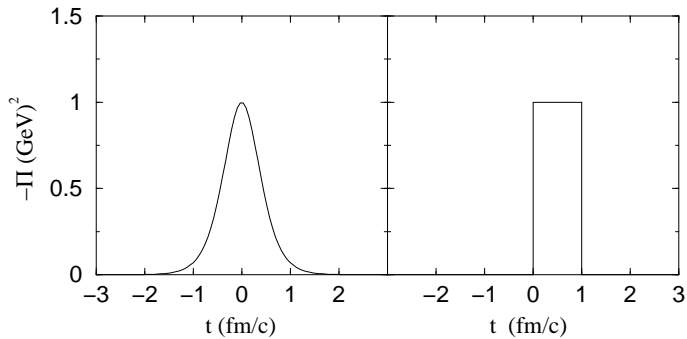}
\end{center}
\caption{Eckart ($|\Pi|=1$ (GeV)$^2$, $T=0.5$ fm/c) and rectangular 
($|\Pi|=1$ (GeV)$^2$, $T=1$ fm/c) perturbations are shown on the 
right and left panels,
respectively.   
\label{potentials}}
\end{figure}
Fig. \ref{n_oned} shows the average number 
of particles produced in the two models described above. 
For our estimate we made the following choice of parameters. 
The effective mass drops to the value of
$m^2_{\rm eff}\,=\,-m_{\sigma}^2/2$, correspondingly the parameter $\Pi$
is chosen at $-1\,\,{\rm GeV}^2$ 
for both  models. The time scale, given by parameter 
$T$, is  1 fm/c and 0.5 fm/c for the square barrier and the Eckart potential, 
respectively. It is important to notice that even though 
the graph is plotted over a continuous variable $k\,,$ our finite size
spatial box allows only 
discrete values of momentum in each  direction. 
With the expected interaction region size of 2 fm, the lowest momentum in a 
cubic well of size $L$
is $|{\bf k}|\approx\sqrt{3}\,\pi/L$, numerically this is about 
540 MeV.  
According to  
Fig. \ref{n_oned}, both  models predict around several 
hundreds of particles. 
The next higher-lying modes,  
have significantly smaller numbers of particles. 
The picture presented  shows that 
one can only hope to have very few modes that actually form 
the condensate, as the number of particles falls drastically for 
higher momenta. This is  consistent with the 
argument that in order to get a noticeable condensate one should 
have energy of the mode dipping below zero. The square barrier 
model provides a simple  estimate for the number of particles produced if 
the mode  just touches zero,
$$\overline{n}\,=\,-{\Pi T^2\over 4}\,\approx\, {m_{\sigma}^2 T^2\over 8}\,
\approx 6 \, ,$$ where the sigma mass $m_{\sigma}$ 
is taken 1.4 GeV and the interaction time  $T$ is around 1 fm/c .  

\begin{figure}
\begin{center}
\epsfxsize=9.0cm \epsfbox{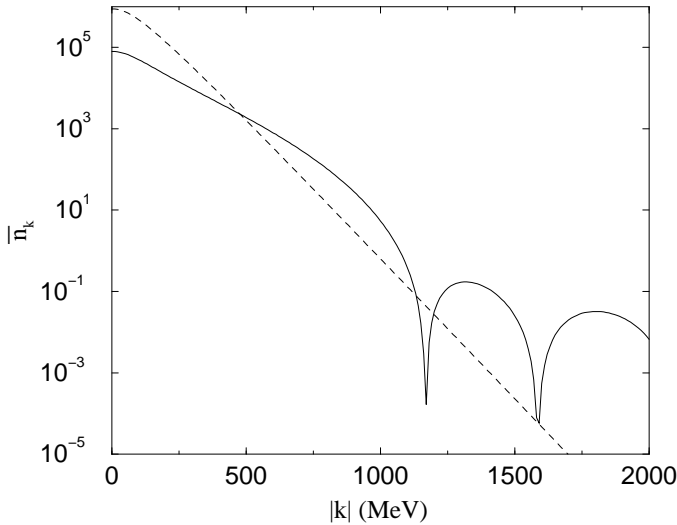}
\end{center}
\caption{The average number of particles produced in
the mode ${\bf k}$ as a function of $k$ for  square potential (solid line)
and Eckart potential (dashed line). The parameters of the perturbation 
are chosen so that $\Pi\,=\, 1\,\, {\rm GeV}^2 $ for both models, and
$T$ is 0.5 fm/c and 1.0 fm/c for Eckart and square potentials, respectively. 
\label{n_oned}}
\end{figure}

For more complicated perturbations it is possible to 
use the WKB approximation in order to determine the reflection coefficient of 
classical  Eq. (\ref{classicaloscillator}) for tunneling.
With the notation
$$\xi\,=\,2\, \exp \left (
\int_{-\infty}^{\infty} |\omega (t)|\, \Theta \left ( 
-\omega^2 (t)\right )\, dt \right )\, 
,$$ where the Heaviside theta-function limits 
the integration to the region of negative $\omega^2(t)\, ,$ we have
the average number of particles given in the semiclassical limit of 
$\xi\gg 1$ by
\begin{equation}
\overline{n}\,={\xi^2\over 4}\,. 
\label{WKB}
\end{equation} 

Concerning the total particle distribution one has to add all particles
from all modes and consider the distribution as a superposition. 
Based on the results of the above examples and  
Sec. 2 we can conclude the following. About a thousand  neutral pions 
with momentum 540 MeV/c are produced by the lowest mode and  are 
distributed according 
to $P_1(2n)$, Eq. (\ref{convolution}). The  distribution over
species is ${\cal P}_1\sim 1/\sqrt{f}\,,$ Eq. (\ref{approxf}).
This is the single-mode result described by Eqs. (\ref{pineutral},
\ref{picharged}, \ref{pitotal}) and  shown in Fig. \ref{oneD}.
Geometry is crucial here, as it determines the energies and
degeneracies of other modes that may or may not compete with the lowest
mode(s).
Higher momentum modes in the box do not ``condense'' if, in  
the scattering picture, their energy is higher than the barrier and reflection
is negligible.
These modes, even
jointly, may on average produce just several particles. The case
of the non-condensate particle production will be considered with a better 
model in the next subsection.  

These simple examples are still far from realistic. One of the major 
failures is that pions from the square box are not real pions and
therefore all excitations that we obtain need to be projected onto
final pion states given by the plane waves of the entire space. 
This projection will produce momentum spread for particles, that will carry the
characteristics of each condensate mode. Nevertheless, these models
produced reasonable results and what is more important they have 
identified the physics of the process.

\subsection{Time- and space-dependent perturbation}
\subsubsection{General solution for a step-like temporal perturbation}
Here we will solve the
perturbed Klein-Gordon equation for the pions in a more 
realistic case, where the effective pion mass is both
space- and time-dependent,  
\begin{equation}
{\partial^2 \pi \over \partial t^2} - \nabla^2 \pi +  
\left ( m_{\pi}^2\,+\,\Pi({\bf x},t)\right ) \pi = 0 \, .
\label{eqmotion1}
\end{equation}
Despite the fact that this dependence is
put into the model by hand, the resulting features can be quite general.
In order to keep our solutions analytic we choose the perturbation
$\Pi({\bf x},t)$ as a step function in time
$$\Pi({\bf x},t)\,=\,\left \{ 
\begin{array}{ll}
\Pi({\bf x}) & {\rm for} \quad t\in[0,T] \\ 
0 & {\rm otherwise}
\end{array}
\right .
\, .$$
This choice will allow us to solve  Eq. (\ref{eqmotion1}) classically
and to construct matrices $u$ and $v$ that control the evolution of a 
plane wave. For most of the discussion the form of $\Pi({\bf x})$
is left as general, but in the last part for the numerical results we 
take it as  a spherical square well.

The wave function $\pi({\bf x},t)$
can be found in each of the time regions, and the solutions should be smoothly 
matched 
keeping $\pi({\bf x},t)$ and the derivative 
$\partial \pi({\bf x},t)/\partial t$ 
continuous. Introducing a separation of variables as
$\pi({\bf x},t)\,={\cal X}({\bf x})\,{\cal T}(t)$ we obtain  equations of
motion for ${\cal X}$ and ${\cal T}$ in all three time regions.
In the  perturbed region  $t\in[0,T]\,,$ Eq. (\ref{eqmotion1}) becomes
\begin{eqnarray}
-\nabla^2 {\cal X}({\bf x})\,+\,\Pi({\bf x})\,{\cal X}({\bf x})\,
=\, {\cal E}\,{\cal X}({\bf x})\,,
\label{perturbedX}
\\
\nonumber
{\partial^2 {\cal T}(t)\over \partial t^2}\,+\, W^2\,{\cal T}(t)\,=\,0\,,
\end{eqnarray} 
where the dispersion relation is 
\begin{equation}
W^2=m_{\pi}^2\,+\,{\cal E}\, .
\label{energy} 
\end{equation}
The unperturbed form of the free space wave equations at $t<0$ or $t>T$
is
\begin{eqnarray}
-\nabla^2 {\cal X}^{\circ}({\bf x})\,
=\, |{\bf k}|^2\,{\cal X}^{\circ}({\bf x})\,,
\label{unperturbedX}
\\
\nonumber
{\partial^2 {\cal T}^{\circ}(t)\over \partial t^2}\,+\, \omega_{\bf k}^2\,
{\cal T}^{\circ}(t)\,=\,0\,,
\end{eqnarray} 
with the usual relation 
$$
\omega_{\bf k}^2=m_{\pi}^2\,+\,|{\bf k}|^2\,. 
$$
Eq. (\ref{unperturbedX})  has  simple plane wave eigensolutions 
that we denote as $|{\bf k}\rangle$ with  positive 
eigenvalues $|{\bf k}|^2\,$. The perturbed 
Eq. (\ref{perturbedX})  with  negative $\Pi$ 
may have negative energy 
bound states as well as the usual continuum states with positive energies,
see Fig. \ref{bound}. For such a bound state, $W^2\,,$ Eq. (\ref{energy}),
becomes negative and $W=i\Omega$ is imaginary.
\begin{figure}
\begin{center}
\epsfxsize=9.0cm \epsfbox{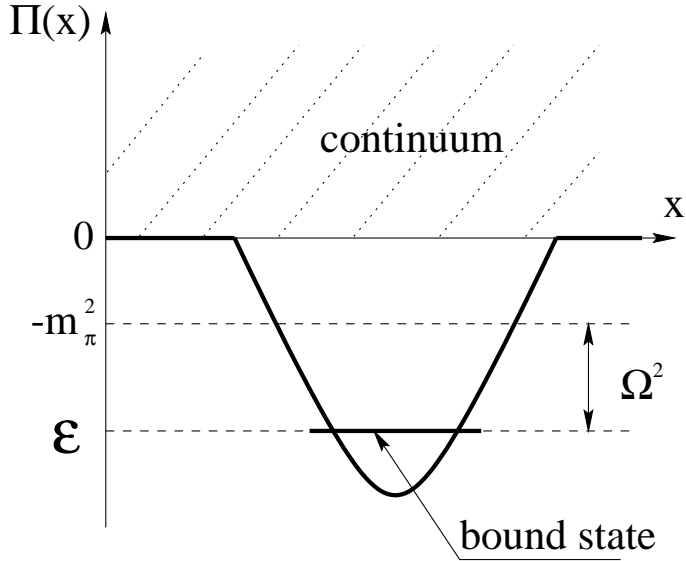}
\end{center}
\caption{The schematic representation of 
the perturbation $\Pi(x)$, with one condensate bound state.
\label{bound}}
\end{figure}
We denote the eigenstates of Eq. (\ref{perturbedX}) as $|\kappa)$ and 
corresponding energy as ${\cal E}(\kappa)\,.$ Here $\kappa$   
is a set of quantum numbers labeling the eigenstates.     
We also assume that both $|{\bf k}\rangle$ and $|\kappa)$ 
are properly normalized and form complete sets,
$$\langle {\bf k}|{\bf k}^{\prime}\rangle\,=\,\delta_{{\bf k},
\,{\bf k}^{\prime}}\,,\quad (\kappa|
\kappa^{\prime})\,=\,\delta_{\kappa,\, \kappa^{\prime}}\,,\quad
1\,=\,\sum_{\bf k}\, |{\bf k} \rangle\,\langle {\bf k}| =
\sum_{\kappa}\, |{\kappa}) \,({\kappa}|  \, .
$$
To determine the Bogoliubov transformation matrices $u$ and $v$ 
we consider the evolution of the wave function 
$$|\pi(t)\rangle \,=
\,e^{-i\omega_{\bf k}t}\, |{\bf k}\rangle\,,\quad {\rm for}\,\,
t<0\,,
$$
through an intermediate stage $t\in[0,T]$ 
$$
|\pi(t)\rangle \,=\,\sum_{\kappa}\, \left ( 
a_{{\bf k}\,\kappa}\,e^{-iW_{\kappa}t}\,+\,b_{{\bf k}\,
\kappa}\,e^{iW_{\kappa}t}
\right )\,|\kappa)\, ,   
$$
into a final state 
$$
|\pi(t)\rangle \,=\,\sum_{{\bf k}^{\prime}} \, \sqrt{\omega_{\bf k} \over
\omega_{{\bf k}^{\prime}}} \, \left ( 
u_{{\bf k}\,{\bf k}^{\prime}}\,e^{-i\omega_{{\bf k}^{\prime}}t}
\,+\,v_{{\bf k}\,{\bf k}^{\prime}}\,e^{i\omega_{{\bf k}^{\prime}}t}
\right )\,|{\bf k}^{\prime}\rangle\, . 
$$
Continuity of the wave function and its derivative with respect to time
allows the determination of unknown coefficients $a_{{\bf k}\,\kappa}$ and
$b_{{\bf k}\,\kappa}$ as well as the matrix elements of interest 
$v_{{\bf k}\,{\bf k}^{\prime}}$ and $u_{{\bf k}\,{\bf k}^{\prime}}\,.$
The result is, both for real and imaginary $W_{\kappa}\,,$
\begin{eqnarray}
\nonumber
u_{{\bf k}\,{\bf k}^{\prime}}\,=\,\sqrt{\omega_{{\bf k}^{\prime}} \over
\omega_{\bf k}} \,{e^{i\omega_{{\bf k}^{\prime}}T}\over 2}
\,\sum_{\kappa}\,\langle {\bf k}^{\prime}|\kappa)(\kappa|{\bf k}
\rangle \times \\ 
\left [ \left (1+{\omega_{\bf k}\over\omega_{{\bf k}^{\prime}} 
}\right )\, \cos \left (W_{\kappa}\,T\right )\,-\, i
\left ({\omega_{\bf k}\over W_{\kappa}}+
{W_{\kappa}\over\omega_{{\bf k}^{\prime}}} \right )\sin \left (
W_{\kappa}\,T\right )
\right ]\,,
\label{uu}
\end{eqnarray}
and 
\begin{eqnarray}
\nonumber
v_{{\bf k}\,{\bf k}^{\prime}}\,=\,\sqrt{\omega_{{\bf k}^{\prime}} \over
\omega_{\bf k}} \,{e^{-i\omega_{{\bf k}^{\prime}}T}\over 2}
\,\sum_{\kappa}\,\langle {\bf k}^{\prime}|\kappa)(\kappa|{\bf k}
\rangle \times \\
\left [ 
\left (1-{\omega_{\bf k}\over\omega_{{\bf k}^{\prime}} 
}\right )\, \cos \left (W_{\kappa}\,T\right )\,-\, i
\left ({\omega_{\bf k}\over W_{\kappa}}-
{W_{\kappa}\over\omega_{{\bf k}^{\prime}}} \right )\sin \left (
W_{\kappa}\,T\right )
\right ]\,.
\label{vv}
\end{eqnarray}
In order to see whether the condensate was 
formed we first address the question of the number of produced pions.
Possible bound states in the solutions of Eq. (\ref{perturbedX}) 
should be carefully treated. As seen below, 
bound states in Eq. 
(\ref{perturbedX}) with the energy ${\cal E}$ below $-m_{\pi}^2$ 
produce an exponential growth of the particle number with time
which is our starting criterion for the search of the condensate.
The average number of particles may be then expressed using Eq. (\ref{number})
as
\begin{equation}
\overline{n}=\sum_{{\bf k}\,,\,{\bf k}^{\prime}}\,
|v_{{\bf k}\,{\bf k}^{\prime}}|^2={1\over 4}
\sum_{\kappa\,,\,\kappa^{\prime}}
\, \left \{ A_{\kappa\,\kappa^{\prime}}\,
\cos\left (W_{\kappa} T\right)\cos \left (W_{\kappa^{\prime}} T\right)+
B_{\kappa\,\kappa^{\prime}}\,
\sin\left (W_{\kappa} T\right)\sin \left (W^{*}_{\kappa^{\prime}} T\right)
\right \}
\,, 
\label{morepions1}
\end{equation}
$$
A_{\kappa\,\kappa^{\prime}}=2\left ( I^{(-1)}_{\kappa\,\,\kappa^{\prime}}
I^{(1)}_{\kappa\,\,\kappa^{\prime}}\,-\,\delta_{\kappa\,\,\kappa^{\prime}}
\right )\,,
$$
$$
B_{\kappa\,\kappa^{\prime}}= \left ({{I^{(1)}_{\kappa\,\,\kappa^{\prime}}}^2\over W_{\kappa}
W^{*}_{\kappa^{\prime}}}-\left({W_{\kappa}\over W^{*}_{\kappa^{\prime}}}+
{W^{*}_{\kappa^{\prime}}\over W_{\kappa}}\right )
\delta_{\kappa\,\,\kappa^{\prime}}+{I^{(-1)}_{\kappa\,\,\kappa^{\prime}}}^2
W_{\kappa} W^{*}_{\kappa^{\prime}} \right )\,,
$$
where we assumed that the states are chosen in such a way that 
the amplitudes 
$\langle {\bf k}| \kappa )$ are real, and we introduced the notations
$$I^{(1)}_{\kappa\,\,\kappa^{\prime}}=I^{(1)}_{\kappa^{\prime}\,\,\kappa} \,=
\,\sum_{{\bf k}}
\,(\kappa |{\bf k} \rangle \omega_{\bf k} \langle {\bf k} |
\kappa^{\prime} ) \, ,
$$
and 
$$I^{(-1)}_{\kappa\,\,\kappa^{\prime}}=I^{(-1)}_{\kappa^{\prime}\,\,\kappa} \,=
\,\sum_{{\bf k}}
\,(\kappa |{\bf k} \rangle {1\over \omega_{\bf k}} \langle {\bf k} |
\kappa^{\prime} )\, .
$$
Depending on a particular level, $W$ may be real or imaginary if 
in Eq. (\ref{energy}) the 
energy ${\cal E}$ is greater or less than $m_{\pi}^2\,;$
nevertheless Eq. (\ref{morepions1}) works for both cases. 

In practice it is  convenient to express the overlap 
of $\langle {\bf k}|\kappa)$ using Eqs. (\ref{perturbedX}, 
\ref{unperturbedX}) for the states  $|\kappa)$ and $|{\bf k}\rangle\,.$
Multiplying Eq. (\ref{perturbedX}) by ${\cal X}^{\circ}$ and 
Eq. (\ref{unperturbedX}) by ${\cal X}$ and subtracting  results we obtain
\begin{equation}
({\cal E}-{\bf k}^2){\cal X}\,{\cal X}^{\circ}\,=\Pi({\bf x})
{\cal X}\,{\cal X}^{\circ}\, + \, \nabla({\cal X}\nabla {\cal X}^{\circ}-
{\cal X}^{\circ}\nabla {\cal X})\, .
\label{current}
\end{equation}
Integration of Eq. (\ref{current}) over all space produces a useful result
\begin{equation}
\langle {\bf k}|\kappa)={1\over {\cal E}-{\bf k}^2}\, \int d^3x\,
\Pi({\bf x}){\cal X}\,{\cal X}^{\circ}\,.
\label{overlap}
\end{equation}
For the continuous spectrum, this expression contains an additional term
with $\delta ({\cal E}-{\bf k}^2) \,.$

Due to  large oscillations of the trigonometric factors in  Eqs. (\ref{vv})
and (\ref{morepions1}),
one should expect considerable particle production only in   
situations with  imaginary $W\,.$
This observation makes it natural to separate the sum in 
Eq. (\ref{morepions1})
into several contributions depending on the intermediate state $|\kappa)\,.$ 
There are exponentially rising terms that
involve transitions through the states with imaginary $W$; it is important 
that the bound states are discrete and their number is finite. 
Another  contribution is of 
all intermediate states that lie in the
continuum. 
This second contribution is always present for any perturbation 
even with no bound states. With a non-zero pion mass there maybe a number
of discrete states that have real $W$ but these states  
have negligible contributions compared to the states from the continuum.
In the remaining part of this subsection we will mainly draw  attention
to the first two cases which we  call ``condensate'' and 
``non-condensate'' pion production.

\subsubsection{Condensate pion production.}
 
In the following picture we assume that there is one state $|\kappa_{0})$
with negative energy so that $W_{\kappa_{0}}=i\Omega\,,$ and  in 
the summation in Eq. (\ref{vv}) 
the term that involves this condensate 
state $|\kappa_{0})$ is dominant.
Thus, the distribution of particles and all other
properties are the same as in the 
single-mode example, i.e. as for the parametrically
excited single oscillator.
According to Eqs. (\ref{number}), (\ref{moments_a}), (\ref{moments}) and 
(\ref{generalcommutaions}),
the traces of  $vv^{\dagger}$ and other higher powers of this matrix 
completely determine all moments of the particle distribution and thus 
the distribution itself. Therefore, it is sufficient to 
consider the eigenvalues of a hermitian
matrix $vv^{\dagger}$.   

Under the assumption that only one term in the sum in Eq. (\ref{vv}) is 
important, we can express the matrix element of $vv^{\dagger}$ as
\begin{equation}
\{vv^{\dagger}\}_{{\bf k},\,{\bf k}^{\prime}}=a {1\over 
\sqrt{\omega_{\bf k}\,\omega_{{\bf k}^{\prime}}}}+
b\,\sqrt{\omega_{\bf k}\,\omega_{{\bf k}^{\prime}}}+c\,\sqrt{
\omega_{\bf k}\over\omega_{{\bf k}^{\prime}}}+
c^{*}\,\sqrt{\omega_{{\bf k}^{\prime}}\over\omega_{\bf k}}\,,
\label{factorizable}
\end{equation}
where the introduced coefficients are
\begin{eqnarray}
a={1\over 4} \left (I^{(1)}\cosh^2 (\Omega T) + \Omega^2 I^{(-1)} 
\sinh^2 (\Omega T) \right )\,,\\
b={1\over 4}
\left ( I^{(-1)} \cosh^2 (\Omega T)+ I^{(1)}\sinh^2 (\Omega T)
/\Omega^2\right )\,,\\
c={1\over 4} \left ( i\left [ I^{(1)}/\Omega+I^{(-1)}\Omega \right ]-1 
\right )\,, 
\end{eqnarray} 
and 
$$ I^{(-1)}\,=\,\sum_{\bf k}\, |\langle {\bf k} | \kappa_{0} )|^2 
{1\over \omega_{\bf k}}
\, , \quad
I^{(1)}\,=\,\sum_{\bf k}\, |\langle {\bf k} | \kappa_{0} )|^2  
\omega_{\bf k}\, .
$$ 
We wrote an element of the matrix $vv^{\dagger}$ in the form of 
Eq. (\ref{factorizable}) in order to show that it consists of 
only four factorizable terms. In accordance with a general theory
of factorizable kernels, the only non-zero eigenvalues of matrix 
$vv^{\dagger}$ are the eigenvalues of the following $4\times4$ matrix
\begin{equation}
\left ( 
\begin{array}{cccc}
a I^{(-1)} & b & c & b I^{(-1)}\cr
a & b I^{(1)} & c I^{(1)} & c^{*} \cr
a I^{(-1)} & b & c & b I^{(-1)}\cr
a & b I^{(1)} & c I^{(1)} & c^{*} \cr
\end{array}
\right ) \quad .
\label{mymatrix}
\end{equation} 
Having the 
second two rows the same as the first two in the matrix  (\ref{mymatrix})
reduces the actual number of non-zero eigenvalues to two, and the secular
equation for the eigenvalues $\lambda$
\begin{equation}
\lambda^2 - \lambda\, {\rm Tr}(vv^{\dagger})+{1\over 16}(I^{(1)}\,I^{(-1)}-1)^2 =0\,.
\label{secular}
\end{equation}
The trace of the matrix $vv^{\dagger}$ in this expression is the same as
the average number of particles produced by one condensate state 
$|\kappa_{0})\,,$
\begin{equation}
\overline{n}={\rm Tr}(vv^{\dagger})=\,{\cosh^2(\Omega\,T)\over 2}\,
\left ( I^{(-1)}\,I^{(1)}\right)
+\,
{\sinh^2(\Omega\,T)\over 4}\,
\left ( {I^{(-1)}}^2\ \Omega^2\,+\,{{I^{(1)}}^2\over \Omega^2}\right )-
{1\over 2}\,.
\label{morepions}
\end{equation}
The result (\ref{morepions}) can be also obtained directly from 
Eq. (\ref{morepions1}).

In the limit of $\exp(2\Omega T)\gg 1$, Eq. (\ref{morepions})
becomes
\begin{equation}
\overline{n}\,=\,{e^{2\Omega T}\over 16}\, \left ( I^{(-1)} 
\Omega\,+\, {I^{(1)}
\over
\Omega}\right )^2 \, .
\label{simpleaverage}
\end{equation}
This approximation is  relevant to our problem as we wish to determine 
the exponentially growing condensate modes and therefore we 
choose a physical environment for which this is true. The limitation
$\exp(2 \Omega T)\gg 1$ is then a criteria for the environment of chiral condensate. 
As seen from Eq. (\ref{secular}), in the same limit  one eigenvalue 
of matrix $vv^{\dagger}$ becomes exponentially large and 
the other one goes to zero. 
This is important because
with only one nonzero eigenvalue we recover the single-mode situation, 
discussed in the previous sections  which leads to the   
distribution
of particles of the form of Eq. (\ref {pineutral}) and a  $1/\sqrt{f}$  
form for the distribution over the pion species. 

As a concluding statement we stress here  that in principle just making the 
approximation  $\exp (2\Omega\, T)\gg 1$ allows us to take one term 
in the sum of Eqs. (\ref{uu}, \ref{vv}) from the  very beginning, and the
same approximation was used again in the end to purify the single mode.
Therefore the condition 
$\exp (2\Omega\, T)\gg 1$ must be a clear indication of a condensate
in the bound mode with relative energy $\Omega\,.$ 
If there are several bound condensate
states with close energies $\Omega$ we would again recover the system
of several modes that was discussed before. 

\subsubsection{Non-condensate pion production}
To make the picture complete we have to  estimate  the 
number of particles that are produced from all the modes 
not involved in the condensate.
The low-lying negative energy states that 
have energies above the pion mass do not make a considerable contribution.
They may be estimated by the same method that was used for a condensate
states just having oscillating exponents instead of exponential growth in
Eq. (\ref{simpleaverage}),
\begin{equation}
\overline{n}\,\sim\,{1\over 16}\, \left ( I^{(-1)} 
\Omega\,+\, {I^{(1)}
\over
\Omega}\right )^2 \, .
\label{simpleaverage2}
\end{equation}
This source does not produce many pions, as will be seen from the numerical
example in the next section.

The contribution to the number of particles from the continuum 
presents a greater
interest. First, as the number of states is infinite, unlike the previous case
of bound states, we may have a significant contribution.
Secondly, it is an important practical question because the continuum states
always produce pions even if the condensate is absent. It  
follows from Sec. \ref{sec_transition}
that the charge distribution of pions from
many modes of almost equal strength 
is Gaussian.
To perform a particle number estimate we again address 
Eq. (\ref{morepions}). It is reasonable to assume that oscillating
terms corresponding to different arguments $W_{\kappa}$ 
and $W_{\kappa^{\prime}}$
average out to zero 
and the contributing terms are those that are in phase, corresponding 
to the same $\kappa\,.$ For these terms 
we take $\sin^2(W T) \approx \cos^2(W T)
\approx 1/2\, .$ Applying these approximations  to 
Eq. (\ref{morepions}) we obtain
\begin{equation}
\overline{n}=\,{1\over 2}\,\sum_{\kappa\,,\,{\bf k}}\,
|\langle {\bf k}|\kappa)|^2 {\left (\omega_{\bf k}-W_{\kappa}\right )^2\over 
\omega_{\bf k}\, W_{\kappa}}\, .
\label{nopions}
\end{equation}
As seen from the structure of this equation, the arbitrary 
$\delta$-contribution in Eq. (\ref{overlap}) does not influence the result
for $\overline{n}\,.$

\subsection{Spatial spherical square well}
The spherical square well potential of a finite depth is the simplest
spatial perturbation $\Pi ({\bf x})$ 
that has an analytical solution. 
We assume that  
$$\Pi({\bf x})\,=\,\left \{ 
\begin{array}{ll}
-V & {\rm for} \quad |{\bf x}|<R \\ 
0 & {\rm otherwise}
\end{array}
\right .
\, .$$
The  rotational symmetry makes  angular 
momentum  a good quantum number, and the remaining  radial dependence 
can be expressed in terms of spherical Bessel functions. 
Due to the largest exponential enhancement the deepest level is expected to
produce most of the contribution. Thus, we concentrate our attention on 
an $s$-wave bound ground state that will be a dominant condensate state 
in the process. By the virtue of symmetry, all pions produced from this
state will have a spherically symmetrical $s$-wave spatial 
distribution. Instead of plane waves it is convenient to quantize spherical
waves
in a large sphere of  radius $L\,.$ 
This leads to the substitution of the former basis  
$|{\bf k}\rangle$ by $|k,l,m\rangle$ where 
$k=|{\bf k}|$ and $l$ and $m$ are the 
orbital momentum and its projection, respectively. 
Needed overlaps between perturbed and non-perturbed states can 
be computed using the one-dimensional version of Eq. (\ref{overlap})
where the 
right hand side is obtained by integrating  Eq. (\ref{current})
from zero up to the size $R$ giving
\begin{equation}
\langle {\bf k}|\kappa)=\left . 
{V\over {\cal E}-{\bf k}^2}\,{({\cal X}(r)\,\partial/
\partial\,r 
{\cal X}^{\circ}(r)-
{\cal X}^{\circ}(r)\,\partial/\partial\,r  {\cal X}(r))\over {\cal E}-k^2+V}
\,\,\right|_{r=R}\,.
\label{squareoverlap}
\end{equation}
In particular, for a bound (${\cal E}<0$) s-state  $|\kappa_{0})$ Eq. (
\ref{squareoverlap}) gives
\begin{equation}
(\kappa_{0} |k, 0,0\rangle={2 \alpha^2 \over \sqrt{L}}\,
{\alpha \cot(\alpha R)\,\sin(k R)
-k \cos (kR)\over \sin (\alpha R)\left (k^2-\alpha^2\right )
\left (k^2+\alpha^2 \cot^2(\alpha R)\right )} \, \sqrt{{\alpha\over
\alpha R-\tan(\alpha R)}}
\label{squareoverlap1}
\end{equation}
where  $\alpha=\sqrt{V-|{\cal E}|}\,.$ 
This overlap is
normalized to one as 
a sum over all momenta $k=\pi n/L\,,$ where $n$ is a nonnegative 
integer. Converting this sum into 
an integral over all positive 
$k$ will remove the quantization radius $L$.
The eigenenergies for  $s$-states are given by the equation
$$\alpha \cot (\alpha R)+\sqrt{{\cal E}}=0\,.$$   
Within these results the matrices $v$ and $u$ can be evaluated 
via Eqs. (\ref{uu}, \ref{vv}) and all the theory described above can be
applied in a straightforward way.

As a realistic physical picture, we take 
the  depth
of the spherical well   $V=1$ GeV$^2$,  radius $R=1$ fm, 
and a lifetime of the condensate  $T=1$ fm/c. Then the lowest level is at 
a depth of ${\cal E}=$(858 MeV)$^2$ which corresponds to 
$\Omega=\sqrt{|{\cal E}|-m_{\pi}^2}= 
847$ MeV. These assumptions are probably  exaggerated as 
the  average number of pions of one particular type in this case is
$\overline{n}\approx2500\, .$ 

Figure \ref{I} displays the behavior of the 
dimensionless variables $I^{(-1)} \Omega$,
$I^{(1)}/\Omega$ and their sum $I^{(-1)} \Omega+I^{(1)}/\Omega$ 
shown in dashed, dotted and 
solid lines, respectively, 
as a function of the potential depth, left panel, and
the potential size, right panel. The singularity of $I^{(1)}/\Omega$ 
at   threshold corresponds to 
$\Omega$ approaching zero and  does not have
any physical significance because the approximation 
$\exp(\Omega T)\gg 1$ does no longer hold.
In general both  plots are dominated by the behavior of 
$\Omega$ for the chosen parameters whereas the 
integrals $I^{(-1)}$ and $I^{(1)}$ are weakly influenced by the form of 
$\Pi({\bf x})$. 
\begin{figure}
\begin{center}
\epsfxsize=9.0cm \epsfbox{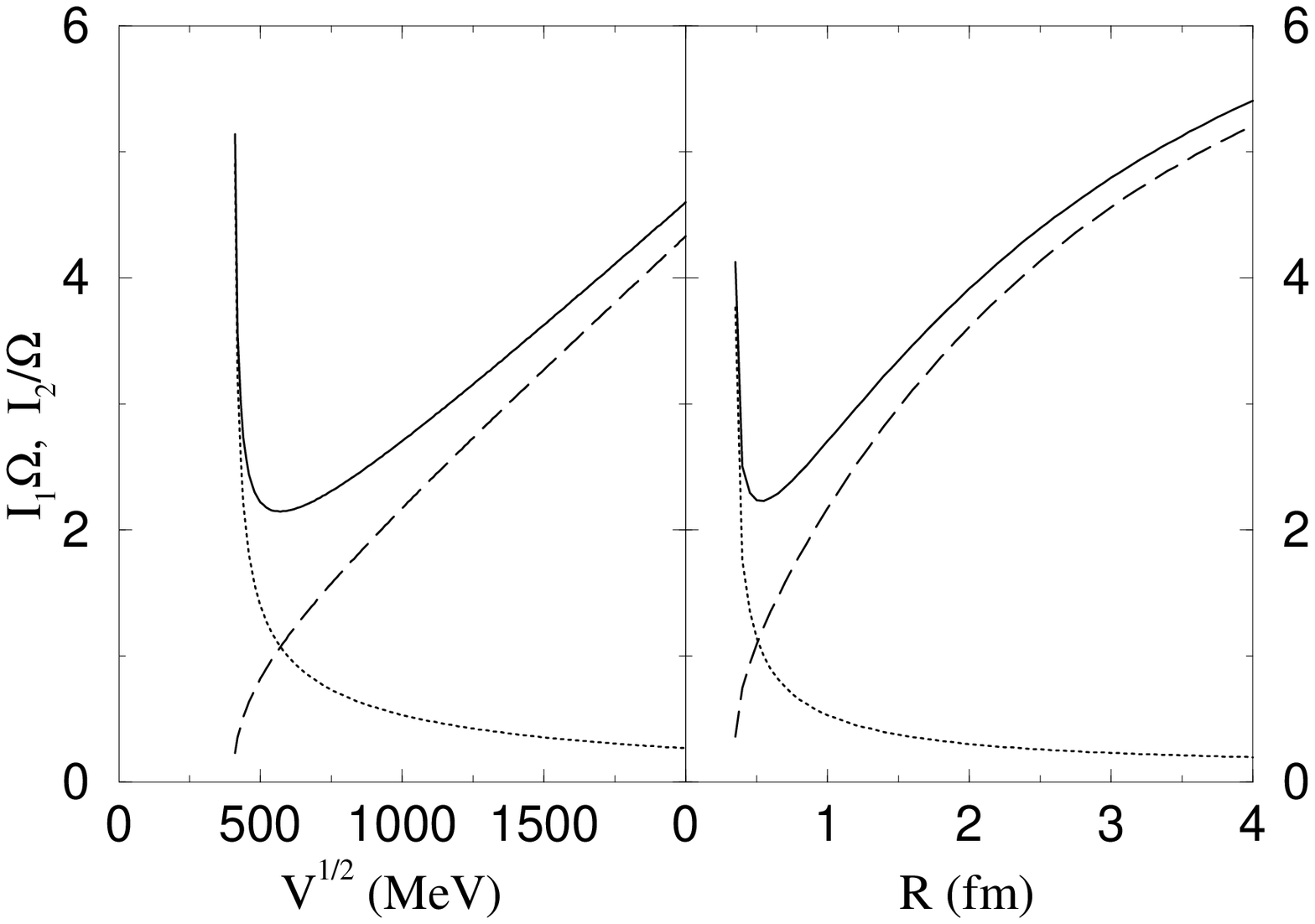}
\end{center}
\caption{The behavior of the dimensionless variables $I^{(-1)} \Omega$,
$I^{(1)}/\Omega$ and $I^{(-1)} \Omega+I^{(1)}/\Omega$ is shown in dashed, dotted and 
solid lines, respectively, as a function of the potential depth $\sqrt{V}$, 
left panel, and potential size $R$, right panel. A fixed size of 1 fm
was used for the left panel and a fixed depth $\sqrt{V}=$1 GeV was
used for the plot in the right panel. 
\label{I}}
\end{figure} 
\begin{figure}
\begin{center}
\epsfxsize=9.0cm \epsfbox{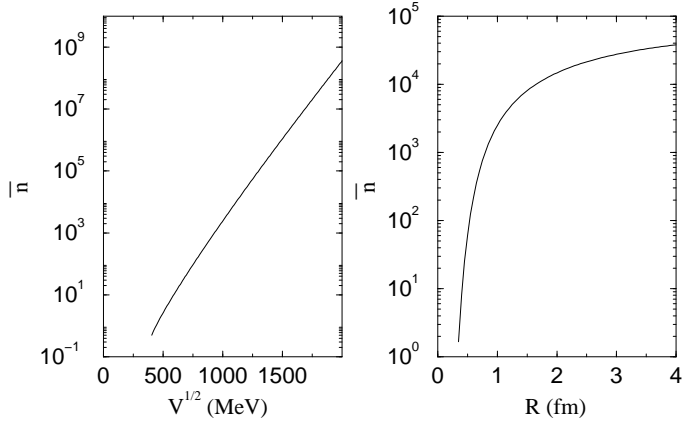}
\end{center}
\caption{The average 
number of particles produced as a function of the depth of the 
potential field $\sqrt{V}$ is shown in the left panel, the size 
$R$ was fixed at 1 fm.
The right panel shows the number of particles produced as a 
function of size $R$ given a fixed depth $V=1$ GeV$^2\,.$
The time length of the perturbation is set at $T=$1 fm/c.
\label{N}}
\end{figure} 

Fig. \ref{N} shows the average number of particles as a
function of the well size, 
right panel, and depth of the perturbation, left panel.
The average number of particles grows approximately  exponentially with
the 
potential depth (Fig. \ref{N}, left panel). This is simply related to the fact 
that in a deep well the 
ground state energy grows almost linearly with the depth having 
$\Omega \approx \sqrt{V}$. By making the potential wider the ground state
approaches the bottom of the well, limiting  $\Omega$  
to a constant. This restrains the growth of particles shown on the 
right panel of Fig. \ref{N}. 
One should bear in mind that the conditions considered  are quite 
extreme and were used here to emphasize the character of the 
condensate. Practically
the time scale may be shorter and perturbation weaker, leading to a
more pre-condensate picture with much fewer pions. The total energy 
available in a heavy ion reaction may provide a guideline to what  the
perturbation $\Pi$ is and  the realistic number
of mesons produced are.  

As a final part of this analysis we consider 
non-condensate pion production, which may be the main mechanism
in most  practical situations.
Eq. (\ref{nopions})
with the additional help of Eq. (\ref{squareoverlap}) 
results in the following form for $s$-wave pions from the continuum
\begin{equation}
\overline{n}={2 V^2 \over \pi^2}\, \int_0^{\infty}d\kappa\,\int_0^{\infty}dk
{\left(k\sin( \alpha R)\, \cos(k R)-\alpha\, \cos(\alpha R) \sin(k R)\right )^2
\over (\alpha^2-k^2)^2 \omega_k W_{\kappa} \left (\omega_k+W_{\kappa} \right
)^2} \, ,
\label{toughintegral}
\end{equation} 
where parameter $\alpha$ is defined as $\alpha=\sqrt{\kappa^2+V}$,  
$W_{\kappa}=\sqrt{\kappa^2+m_{\pi}^2}\,,$ and 
$\omega_k=\sqrt{k^2+m_{\pi}^2}$ are the 
total energies of the corresponding
modes.
As expected, the number of particles 
produced with no condensate involved is quite 
small. The left panel in 
Fig. \ref{nocondensate} shows the average number of non-condensate
pions as a function of the potential depth for  different spatial 
sizes $R=$0.5, 1, 2 fm. The right panel of the same figure displays
the dependence on the size $R$ for various values of $V\,.$ 
The important conclusion here is that the number of non-condensate
pions ranges from a few up to maybe a dozen for extreme cases. 
This number is completely negligible in the presence of a strong condensate
that produces hundreds of mesons. However, present experiments may 
just barely reach the point of the phase transition, and therefore the fraction
of non-condensate pions is considerable, if not dominant. Moreover, other
(conventional) mechanisms of pion production have to be taken into account. 
\begin{figure}
\begin{center}
\epsfxsize=12.0cm \epsfbox{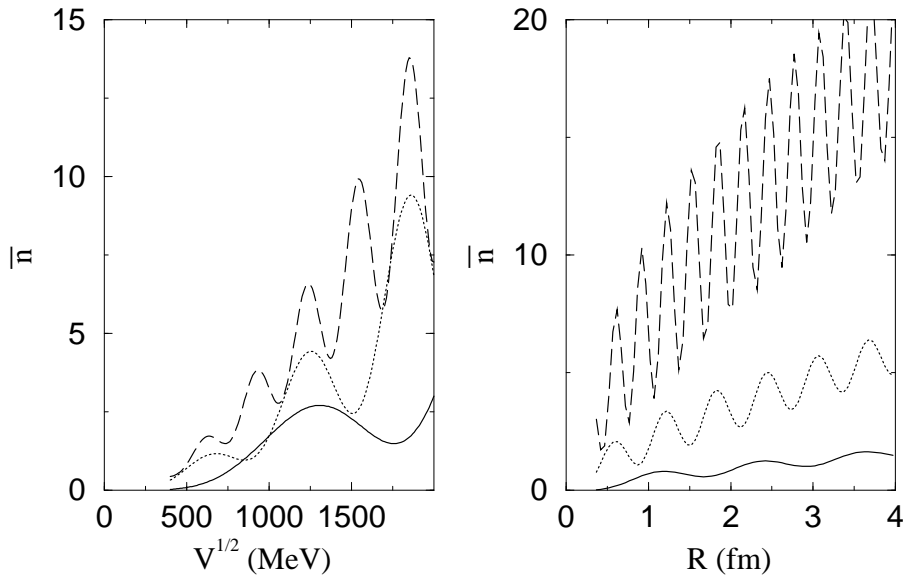}
\end{center}
\caption{
The left side shows the average number of pions of a particular type
as a function of $V\,,$ the depth of the  perturbation. 
Curves displayed as 
solid, dotted and dashed lines correspond to the values of the radius $R$
of 0.5, 1, and 2 fm, respectively. 
Plotted on the right hand side is 
the number of pions versus the radius $R$ for values of 
$\sqrt{V}$ of 0.5, 1 and 2 GeV as solid, dotted and dashed lines, respectively.
\label{nocondensate}}
\end{figure} 

\section{Summary and conclusions }
\label{sec_conclusions}
Our prime objective in this work was to study  a mechanism of  
pion production
in heavy ion collisions related to the creation of a chiral condensate
and to explore how pion distributions can
signal the presence of a condensate. We studied meson 
production by imposing
a pion dispersion relation specific to the medium, i.e. with 
a  space- and time-dependent effective mass. 

In general, the problem of parametric excitation of the field quanta 
presents an interesting question  as it is encountered 
in many branches of physics  from condensed matter to high 
energy physics. The problem also exhibits a vast variety of solutions
ranging from adiabatic  to phase transitions and condensates.

We have conducted an extensive study of 
quantum field equations of a general form of Eq. (\ref{fieldequation}). 
In our picture the parametric excitation of the field 
quanta is carried out 
by the externally given space- and time-dependent mass term.
This term in the Lagrangian  is quadratic in the field,
and the states produced are often called ``squeezed'' 
\cite{picinbono70,zhang90,difilippo92,aliaga93}. 
Some analogy can
be drawn here from  well studied linear current type terms that produce
coherent states. 
However, the essential difference between coherent and squeezed states 
arises due to the fact that the latter correspond to the pairwise generation
of quanta. In the case of pseudoscalar pions, this means that charge,
isospin and parity are exactly preserved.
We have emphasized the fact that the quantum solution
can be built from the classical solution, and this important link
was established via a canonical Bogoliubov transformation. 

With our interest lying in the direction of chiral condensate we
focused our attention on the potential of Eq. (\ref{fieldequation}) 
to form a condensate with fast particle production and
large correlations when the
effective mass goes through zero. 
Along with a general formalism that can be used for numerical
studies we have  analytically solved the problem when the effective mass
experiences sudden abrupt changes.
Consideration of this particular temporal 
perturbation allowed us to clearly separate
the exponentially rising collective pion condensate modes for any 
given spatial form of the perturbation in the effective mass.  
This  produced  conditions where the condensate and its signatures 
can be seen.

We have identified two basic channels of pion production.
The first  involves only a few discrete condensate modes and a large 
associated pion population.
The second leads to the production of far fewer (non-condensate) particles
with a broad phase space distribution.  
Mathematically these channels can be identified as production of 
mesons from bound and  continuum states 
of a Schr\"odinger equation with a 
potential of the form of the perturbation itself, see Fig. \ref{bound}.
The  bound 
states of negative energy are responsible for the characteristic 
features of the condensate.

Numerically, the number of non-condensed pions ranges from a few up to a 
dozen, and as expected is not very sensitive to the choice of the
spatial and temporal form of the perturbation in the effective mass. 
In contrast, 
the condensate modes have an exponential 
sensitivity to the input parameters. 
As the abrupt changes in the effective pion mass grow in strength, a
critical point is reached with the appearance of the condensate mode.
The population of this mode increases dramatically  from zero to 
thousands with a further slight change in the mass parameter.
Due to this hyper-sensitivity of the number of condensed pions to the 
perturbation it is 
practically impossible to predict the effect quantitatively 
without specifying precisely the scenario of the process. 

However, our results predict the number of non-condensate pions, thereby
imposing a lower limit on the statistics needed to unambiguously detect 
the chiral condensate.
Furthermore, we have shown that the condensate pions have a 
specific momentum distribution due to their common collective 
condensate mode. We have also shown that 
although the distribution over species starts 
from the famous  $1/\sqrt{f}$ form 
for one mode it quickly becomes Gaussian with the appearance of successive
modes. Therefore the presence  of several modes complicates the
detection of a chiral condensate. 
The number of modes present increases with energy. 
In addition, the  mass parameter can be strongly perturbed in
more than one region, each increasing the number of modes. 
Tunneling and chaotic dynamics 
in the resulting multi-well potential might lead to another class of 
interesting problems.

This work can be extended in several directions.  
With the formalism presented here,  large numerical studies
of the effective pionic field in a hot medium can be conducted 
involving realistic and even self-consistent forms of  the perturbation 
given by the $\sigma$ field.  Constraints on the perturbation
of the mass parameter should be related  more rigorously to observables.  
Further analysis of our results applied to
phase transitions, zero mass particle production, energy transfer 
and many other field theory problems would definitely be fruitful. 
We feel that this work may provide a step forward in the study 
and classification
of  field theories with parametric excitations and possibly clarify the
nature of the produced squeezed states.

\acknowledgments{The authors acknowledge support from NSF Grant 96-05207.}

\end{document}